\documentclass[prd,aps,10pt,floats,floatfix,nofootinbib]{revtex4-1}
\pdfoutput=1
\usepackage[utf8]{inputenc}
\usepackage{amsmath}
\usepackage{amsfonts}
\usepackage{amssymb}
\usepackage{amsthm}
\usepackage{hyperref}
\usepackage{xcolor}
\usepackage{bm}
\usepackage{graphicx}
\usepackage{feynmp-auto}
\usepackage{mathrsfs}
\usepackage{caption}
\usepackage{subcaption}
\usepackage{framed}

 \def\be{\begin{equation}}
\def\ee{\end{equation}}
 \def\ba{\begin{align}}
\def\ea{\end{align}}
\def\bea{\begin{eqnarray}}
\def\eea{\end{eqnarray}}

\def\a{\alpha}
\def\b{\beta}

\def\m{\mu}
\def\n{\nu}

\def \Horava {Ho\v rava }

\newcommand{\bseq}{\begin{subequations}}
\newcommand{\eseq}{\end{subequations}}

\begin{document}
\title{{\bf The Status of Ho\v rava Gravity}}
\author{M. Herrero-Valea}
\email[]{mherrero@ifae.es}
\address{Institut de Fisica d’Altes Energies (IFAE), The Barcelona Institute of Science and Technology, Campus UAB, 08193 Bellaterra (Barcelona), Spain}

\begin{abstract}
Ho\v rava gravity is a proposal for a UV completion of gravitation obtained by endowing the space-time manifold with a preferred foliation in space-like hypersurfaces. This allows for a power-counting renormalizable theory free of ghosts, at the cost of breaking local Lorentz invariance and diffeomorphism invariance down to foliation preserving transformations. In this updated review, we report the main successes and challenges of the proposal, discussing the main features of the projectable and non-projectable versions of Ho\v rava gravity. We focus in three main aspects: (i) the UV regime, discussing the renormalizability and renormalization group flow of the projectable theory, as well as the obstacles towards similar results in the non-projectable case; (ii) the low energy phenomenology of both models, including the PN regime, the most updated constraints in the parameter space of the theory, the structure of black holes at low energies, and the possibility of dark matter emerging from gravitational dynamics in the projectable model; and (iii) the specific phenomena induced by higher derivatives, such as the possibility of regularizing singularities, the dynamical behavior of solutions to dispersive equations, and the emission of Hawking radiation by universal horizons.
\end{abstract}

\maketitle
\tableofcontents
\newpage

\section{Introduction}

More than a century after its formulation, General Relativity (GR) has been tested down to an exquisite level at large scales and low energies. Examples of this include sub-milimeter experiments, which verify the validity of Newton's law at those scales \cite{Hoyle:2004cw}; as well as Solar system experiments, which are able to measure several orders of post-Newtonian (PN) effects  \cite{Will:2014kxa} — corrections in powers of $v/c$, where $v$ is the characteristic speed of a system and $c$ is the speed of light. These tests probe the weak limit of gravitation, but investigations in the strong field regime have also been put forward \cite{will_2018} -- precise measurements of the period of binary systems, in particular of binary pulsars, where one component is a millisecond pulsar; as well as the direct observation of gravitational waves (GW) from the merger of compact objects \cite{Berti:2015itd,Barack:2018yly}. These allow to confirm the predictions of GR in environments with large potentials and large velocities. The predictions of Einstein's theory have also been extensively confirmed in Cosmology, where the cosmological standard model $\Lambda\text{CDM}$ reproduces observations with an astonishing precision \cite{Planck:2018vyg}.

In spite of this agreement, all these tests only probe long distances, and the extrapolation of GR predictions below sub-atomic scales encounters obstructions. The most well-known of them is the presence of singularities in black hole geometries and cosmological solutions, which demands an understanding of gravitational dynamics in environments with high curvatures and energies, where quantum gravitational effects presumably become important \cite{Burgess:2003jk}. However, the inclusion of GR within the framework of Quantum Field theory (QFT) leads to a non-renormalizable theory, due to the dimensionful character of the Newton's constant $G$, which scales as an inverse mass squared. Every loop in the perturbative expansion brings down new divergences which require new counter-terms, schematically ordered as a power series in $G \nabla^2$, where $\nabla$ denotes a derivative acting on a metric. As such, the perturbative expansion of the effective action takes the form of an Effective Field Theory (EFT) expansion \cite{Donoghue:1994dn}, with cut-off $\Lambda \sim G^{-\frac{1}{2}}$. This leads to conclude that the Einstein-Hilbert action is only the first term in the low energy expansion of an unknown theory of Quantum Gravity. Predictions of GR, and of its QFT version obtained by following EFT methods, necessarily break down before an energy scale $E\sim \Lambda$, and the search of an ultra-violet (UV) completion of gravitational interactions has become by now one of the most important endeavours in modern theoretical physics.

As we have mentioned, the problem of the non-renormalizability of GR can be isolated in the dimensionful character of the Newton's constant. Due to this, interaction vertices for excitations of the gravitational field around flat space -- after proper normalization -- scale with positive powers of the dimensionless quantity $G p^2$, where $p$ is the momentum of the perturbation \cite{DeWitt:1967ub}. As the number of vertices increases, the decay of the propagator with $p^{-2}$ is not enough to force all diagrams to converge or diverge at most logarithmically \cite{Goroff:1985th}. This leads eventually to a tower of new divergences entering with every new loop order. A simple solution to this problem can be found by extending the Lagrangian with terms quadratic in the curvature, so that the propagator changes and decays at large energies as $p^{-4}$. This leads to Quadratic Gravity \cite{Stelle:1977ry,Salvio:2018crh} and it is indeed enough to render the theory renormalizable, as proven long ago by Stelle \cite{Stelle:1976gc}. However, this also introduces an extra propagating degree of freedom in the spectrum of the theory, a massive graviton with mass of order $\sim \Lambda$, whose kinetic term has the wrong sign and behaves as an Orstrogradsky ghost, compromising unitarity. Despite large efforts in the recent years to make sense, or get rid, of this ghost graviton, no definitive answer exists yet \cite{Salvio:2018crh}. 

The presence of the ghost in Quadratic Gravity is intimately tied to the fact that higher derivative terms in the action introduce higher time derivatives \cite{Woodard:2015zca}, which inevitably lead to the existence of the ghost unless the system is degenerated\footnote{It is actually possible to write higher derivative theories without ghosts for scalar-tensor theories, thanks to this subtlety. These are known as Degenerate Higher Order Scalar-Tensor (DHOST) theories \cite{Langlois:2015cwa,Langlois:2017mdk}. However, the requirement of diffeomorphism invariance of the action obstructs a similar construction in pure gravity. The massive ghost is always present whenever the action is renormalizable.}. A possible way out of this is to include only higher order \emph{spatial derivatives}, while keeping the action second order in time derivatives, but this will immediately break both diffeomorphism and local Lorentz invariance in an explicit way. Nevertheless, this is the key idea behind the formulation of \Horava Gravity \cite{Horava:2009uw}, which UV completes GR by allowing for this explicit breaking of diffeomorphisms down to only those transformations that preserve a preferred time direction, defined in terms of a foliation into co-dimension one spatial surfaces. This permits to keep the theory second order in time derivatives explicitly for all observers. By doing this, we can achieve a theory which is explicitly power-counting renormalizable.

Ho\v rava's proposal quickly boosted a large number of research works, testing the consistency of the theory and examining its predictions. It was quickly shown that the original proposal had problems due to extra conditions imposed to reduce the number of possible terms in the Lagrangian \cite{Blas:2009yd,Blas:2009qj}, which in a general formulation can be of ${\cal O}(50)$. In order to alleviate this, \Horava introduced two simplifications -- an extra symmetry in terms of detailed balance, and the condition of projectability of the foliation. However, both of them lead to issues when contrasted with Nature. Detailed balance, regardless of whether projectability is also enforced, implies a non-vanishing cosmological constant of the wrong sign, and includes a term which violates parity. This automatically rules out the model. Projectable \Horava gravity (pHG) without detailed balance, on the other hand, enters into strong coupling at low energies, but it has seen many successes despite this. It has been shown to be renormalizable beyond power counting \cite{Barvinsky:2015kil}, so that all divergences correspond to gauge invariant operators in the bare Lagrangian, and displays asymptotically free fixed points both in $2+1$ \cite{Barvinsky:2017kob} and $3+1$ dimensions \cite{Barvinsky:2021ubv}. It also leads naturally to a Dark Matter (DM) contribution to the cosmological matter density without the need of extra ingredients \cite{Mukohyama:2009mz}. The problem of strong coupling has also been discussed in analogy to confinement in Quantum Chromodynamics (QCD) \cite{Blas:2018flu}.

Nevertheless, the most general action for \Horava Gravity is that of \emph{non-projectable \Horava Gravity} \cite{Blas:2009qj}, obtained by allowing all possible terms in the Lagrangian, without imposing projectability. This is the theory that we will refer to as \Horava Gravity. Although it contains a very large number of terms, these solve the strong coupling of the projectable version. However, abandoning projectability also obstructs a direct proof of renormalizability along the lines of \cite{Barvinsky:2015kil}, due to the symmetry structure of the theory. The presence of extra terms in the action also leads to Lorentz violations at all energies \cite{Jacobson:2013xta}, not only in the UV regime. This is potentially dangerous if they are very large, but it also provides ways to probe the theory with low energy tests and observations, similar to those carried out in GR \cite{Will:2014kxa,Bettoni:2017lxf,Blas:2012vn}. In particular, observations of GW emission from compact objects leads to constraints which are competitive with those in Einstein's theory \cite{Yagi:2013ava,Yagi:2013qpa,Gupta:2021vdj}. Actually, even with this percolation of Lorentz violations down to the infra-red (IR), the low energy phenomenology of \Horava Gravity is surprisingly close to that of GR. Solutions with large symmetry groups, such as spherical or axially symmetric, and Friedmann–Lema\v itre–Robertson–Walker (FLRW) space-times, lead to similar observational results, even though they exhibit different theoretical properties. Hawking radiation in black hole space-times is also present, with striking similarities to analogue gravitational systems \cite{Herrero-Valea:2020fqa}.

All these results shape \Horava Gravity as one of the few -- if not the only -- modified gravity theories that ticks several boxes at the same time. In particular, it has not been ruled out by GW observations, reproduces all the most important milestones of GR -- such as the existence of black holes, and accelerated expanding large scale solutions --, and provides a UV completion for gravitational interactions with no ghosts and power-counting renormalizability. However, there are still many dark spots where to shed light. Full renormalizability of the non-projectable theory stands out as a long-standing question, which remains unanswered despite recent advances. An assessment of phenomenological implications of higher derivative terms in the action -- such as stability of black hole solutions -- is also missing. The understanding of pHG as a realistic theory á la QCD, by studying its strongly coupled region, is also an open problem.

In this review, we will try to offer a detailed glimpse over all these points, discussing the status of the different versions of \Horava Gravity at the time of writing this text, and giving an overview of future lines of research and open questions. Notice that other review works have been written in the past on this topic \cite{Mukohyama:2010xz,Sotiriou:2010wn,Wang:2017brl,Barvinsky:2023mrv}. This text aims to extend and complement them in order to shape a global picture of the status of the research in Ho\v rava gravity. 

With this goal, this manuscript is organized in the following way. We start by introducing Ho\v rava gravity in Section \ref{horava_grav}, describing the projectable and non-projectable versions in subsections \ref{sec:pHG} and \ref{sec:npHG}. Later, we focus on the UV regime of the theory, describing the algebra of constraints (subsection \ref{sec:hamiltonian}), the renormalization of the projectable model (\ref{sec:renorm_pHG}), and the challenges towards renormalizing the non-projectable theory (\ref{sec:renorm_nphg}). In Section \ref{sec:low_energies} we discuss several aspects of the low energy phenomenology of Ho\v rava gravity -- its connection to Einstein-Aether gravity (\ref{sec:EA_k}), the constraints in the parameter space obtained through a parameterized post-Newtonian expansion (subsection \ref{sec:constraints}), black hole solutions at low energies (\ref{sec:black_holes}), the possible percolation of Lorentz violations to Standard Model species (\ref{sec:percolation}), and the possibility of dark matter emerging as a consequence of Lorentz violations (\ref{sec:sec:DM_integration}). Finally, we will explore some explicit consequences of the higher derivative operators that characterize Ho\v rava gravity in Section \ref{sec:beyond}. We will first discuss the possibility of UV completing black hole geometries in subsection \ref{sec:UV_BHs}, and later move to the exploration of the dynamics of a scalar field sharing some of its properties, in subsection \ref{sec:toy}. We will also show how Hawking radiation can emerge from universal horizons in subsection \ref{sec:hawking}. We will draw conclusions and discuss possible future research directions in Section \ref{sec:conclusions}.


\section{Ho\v rava Gravity in a nut-shell}\label{horava_grav}
General Relativity is formulated as a field theory in a $d$-dimensional manifold ${\cal M}$ equipped with a metric $g_{\m\n}$ and inariant under diffeomorphisms, with action
\begin{align}\label{eq:E-H}
    S_{\rm GR}=-\frac{1}{16\pi G}\int d^{d}x \sqrt{|g|} \ (\hat{R}-2\hat \Lambda)+\int d^{d}x \sqrt{|g|}\ {\cal L}_{\rm M},
\end{align}
where $\hat{R}$ is the scalar curvature, $\hat \Lambda$ the cosmological constant, $g=\det(g_{\m\n})$, and ${\cal L}_{\rm M}$ denotes a generic matter Lagrangian. The Newton's constant $G$ has energy dimension $[G]=2-d$, which corresponds to $[G]=-2$ in four space-time dimensions. Variation of \eqref{eq:E-H} with respect to the metric leads to the Einstein equations
\begin{align*}
    \hat{R}_{\m\n}-\frac{1}{2}g_{\m\n}\hat{R}+\hat{\Lambda}g_{\m\n}=8\pi G T_{\m\n},
\end{align*}
where $T_{\m\n}=-\frac{2}{\sqrt{|g|}}\frac{\delta}{\delta g^{\m\n}}\left(\sqrt{|g|}\ {\cal L}_{\rm M}\right)$ is the energy-momentum tensor of ${\cal L}_{\rm M}$.

In absence of sources, and when $\hat \Lambda=0$, the natural vacuum solution is $g_{\m\n}=\eta_{\m\n}$, where $\eta=\text{diag}(1,-1,\dots,-1)$ in Cartesian coordinates\footnote{Hereinforward we use mostly minus signature.}. Perturbations around this solution correspond to $g_{\m\n}=\eta_{\m\n}+\sqrt{16 \pi G}\ h_{\m\n}$, where the factor $16 \pi G$ ensures that the kinetic term of the graviton $h_{\m\n}$ is properly normalized. Expanding \eqref{eq:E-H} to quadratic order and appending it with a gauge fixing term to eliminate the redundancy introduced by linearised diffeomorphisms
\begin{align}\label{eq:dedonder}
    S_{gf}=-\frac{1}{16\pi G}\int d^dx\ F_\m F^\m,\quad F^\m=\partial_{\n}h^{\m\n}-\partial^\m h,
\end{align}    
we can obtain the propagator of the graviton field $h_{\m\n}$ in a straightforward way
\begin{align}
    \langle h_{\m\n}(p)h_{\a\b}(-p)\rangle=\left(\eta_{\m\a}\eta_{\n\b}+\eta_{\m\b}\eta_{\n\a}-\eta_{\m\n}\eta_{\a\b}\right)\frac{i}{p^2+i\epsilon},
\end{align}
where we have used the standard $i\epsilon$ prescription.

Interactions start at cubic order in the graviton perturbation, and are derivatively coupled, leading to momentum dependence in interaction vertices \cite{DeWitt:1967ub}. The normalization of $h_{\m\n}$ introduces positive powers of $G$. A typical cubic term takes the schematic form
\begin{align}
    \langle h_{\m\n}(p_1)h_{\a\b}(p_2)h_{\rho\sigma}(p_3)\rangle= \sqrt{G}\ q^2 \times {\cal T}_{\m\n\a\b\rho\sigma}(\eta,\hat p_1,\hat p_2, \hat p_3)\times \delta^d(p_1+p_2+p_3),
\end{align}
where $q$ can be either $p_1$, $p_2$ or $p_3$, $\hat p_i=p_i/|p_i|$, and ${\cal T}$ denotes an unspecified tensor structure with the appropiate symmetries. Note that the scaling of a vertex with $G$ grows with the number of legs -- a generic vertex with $l$ legs will scale as $G^{\frac{l-2}{2}} q^2$. Due to this, the superficial degree of divergence of a generic diagram with $L$ loops behaves as
\begin{align}
    {\rm div}=(d-2)L+2=2-[G]L.
\end{align}
As we can see, this grows with the number of loops, unless $d=2$, in which case $[G]=0$ and GR becomes a renormalizable -- and also topological -- theory \cite{Polyakov:1987zb}. For every other $d>2$ the theory is non-renormalizable, generating new divergences at every loop order and requiring counterterms with higher powers of momentum. The out-coming effective action from a perturbative expansion around an arbitrary background $g_{\m\n}$ becomes a power series in the Riemann tensor, whose explicit form depends on $d$ \cite{Donoghue:1994dn,Barvinsky:1985an}.

A possible way out of this issue in $d=4$ is to include terms quadratic in the curvature -- $R_{\m\n\a\b}R^{\m\n\a\b}, R_{\m\n}R^{\m\n},$ and $R^2$ -- in the bare action, ending up with the action of Quadratic Gravity \cite{Salvio:2018crh,Stelle:1977ry}. These not only yield new interaction vertices proportional to $G p^4$, but also modify the propagator of the gravitational field. It can be seen from the schematic re-summation
\begin{align}\label{eq:prop_4}
    \frac{1}{p^2}+\frac{1}{p^2}G p^4 \frac{1}{p^2}+\frac{1}{p^2}Gp^4 \frac{1}{p^2}Gp^4\frac{1}{p^2}+\dots =\frac{1}{p^2 -G p^4},
\end{align}
so that the UV limit of the propagator scales as $p^{-4}$. This modifies the superficial degree of divergence to ${\rm div}=(d-4)L+4-2V_2$, where $V_2$ is the number of vertices which scale as $p^2$, coming from the Einstein-Hilbert term. In contrast to before, for $d=4$ the degree of divergence is now independent of the number of loops and remains bounded by four, which is the maximum power of momentum in the bare Lagrangian. This means that all divergences can be absorbed in quantities with up to four derivatives. Hence, the theory is power-counting renormalizable\footnote{Full renormalizability also requires all divergences to be proportional to gauge-invariant operators. A classical proof for Quadratic Gravity can be found in \cite{Stelle:1976gc}}. 

In spite of this, the corrected propagator \eqref{eq:prop_4} entails problems, since it can be decomposed as
\begin{align}
    \frac{1}{p^2 -G p^4}=\frac{1}{p^2}-\frac{1}{p^2-G^{-1}},
\end{align}
which signals that together with the massless graviton, Quadratic Gravity also propagates a massive graviton\footnote{A precise computation shows that Quadratic Gravity propagates also a degree of freedom in the scalar sector. The later is controlled by the operator $R^2$ in the action, while the massive graviton is introduced through $R_{\m\n}R^{\m\n}$. Excluding this last operator still leads to a fourth derivative theory, which can be however written as a second order scalar-tensor theory, and remains non-renormalizable.}, with mass $m\sim G^{-\frac{1}{2}}$ in $d=4$. The sign in the residue of the propagator however has the wrong sign, signaling that this degree of freedom behaves as an Orstrogradsky ghost \cite{Woodard:2015zca}. Its Hamiltonian is not bounded by below, which leads to classical runaway solutions and loss of unitarity at the quantum level. There are several attempts to get rid of this ghost mode in the recent literature, but there has not been clear success so far. For a review on the topic, see \cite{Salvio:2018crh}.

Note that the previous problem is connected to the presence of higher \emph{time} derivatives in the Lagrangian. Any non-degenerate kinetic term with more than two time derivatives inevitably leads to the presence of ghost fields. An obvious solution is then to avoid the introduction of higher time derivatives, regularizing the high-energy limit of the theory by spatial derivatives only. Although this automatically breaks local Lorentz invariance of the action, it does the work. Adding terms with the schematic form\footnote{From now on we will use greek letters for $d$-dimensional space-time indices, while latin letters will run over spatial directions only.} ${\cal G}(\partial_i)^{2z}$, where ${\cal G}$ is a coupling constant and $z>1$, we can repeat the argument in \eqref{eq:prop_4} and arrive to
\begin{align}\label{eq:prop_lifshitz}
    \frac{1}{p^2}+\frac{1}{p^2}{\cal G}(k_i)^{2z} \frac{1}{p^2}+\frac{1}{p^2}{\cal G}(k_i)^{2z} \frac{1}{p^2}{\cal G}(k_i)^{2z}\frac{1}{p^2}+\dots =\frac{1}{p^2 -{\cal G}(k_i)^{2z}},
\end{align}
where we have decomposed the space-time momentum as $p^\m=(\omega,-k^i)$. For a sufficiently large value of $z$ this leads to a renormalizable theory which, in contrast to the case of Quadratic Gravity, remains second order in time derivatives, avoiding the presence of Orstrogradsky ghosts. Instead, the physical pole of the propagator develops a modified dispersion relation $\omega^2=k^2+{\cal G}k^{2z}$. At energies below a scale $\Lambda^{2-2z}\sim {\cal G}$ the dispersion relation is dominated by the $k^2$ term and the addition of the new operator can be seen as a small deformation, so that Lorentz invariance seems to emerge naturally at low energies.

\Horava Gravity \cite{Horava:2009uw} aims to reproduce this behavior by combining two main ingredients. First, we append the manifold ${\cal M}$ with a foliation structure ${\cal F}$ in co-dimension one hypersurfaces \cite{Gourgoulhon:2007ue}, such that the normal vector to the foliation $U^\m$ defines a universal time coordinate $t$. This reduces the symmetry of the theory down to foliation preserving diffeomorphisms (FDiff) \cite{Blas:2010hb}
\begin{align}\label{eq:fdiff}
    t\rightarrow t'(t),\quad x^i\rightarrow x'^i(t,x),
\end{align}
where $t'(t)$ is a monotonous function and $x^i$ are the coordinates in the spatial leafs of the foliation. The second ingredient is an anisotropic scaling symmetry of the Lifshitz kind \cite{Hohenberg:1977ym,Ardonne:2003wa} in the UV regime
\begin{align}\label{eq:lifshitz_scaling}
t\rightarrow \chi^{-z} t,\quad x^i\rightarrow \chi^{-1} x^i,
\end{align}
which defines power-counting at high energies, with $\chi$ constant. For $z=1$ we recover the usual scaling of relativistic quantum field theories (QFTs), while $z>1$ breaks Lorentz invariance explicitly. From now on we will refer to the dimension attached to this as \emph{scaling dimension}, denoting it with a subscript $s$, e.g. $[t]_S=-z$.

In the presence of a foliation, it is then natural to adopt coordinates following Eulerian observers, by writing the metric in the Arnowitt, Deser, and Misner (ADM) form \cite{Arnowitt:1959ah}
\begin{align}
    ds^2=N^2 dt^2-\gamma_{ij}(dx^i+N^i dt)(dx^j+N^j dt),
\end{align}
where $N, N^i$ and $\gamma_{ij}$ are the lapse, shift and metric of the spatial leafs, respectively. Under \eqref{eq:fdiff} they transform as
\begin{align}
    N\rightarrow N\frac{dt}{dt'},\quad N^i\rightarrow \left(N^j \frac{\partial x'^i}{\partial x^j}-\frac{\partial x'^i}{\partial t}\right)\frac{dt}{dt'},\quad \gamma_{ij}\rightarrow \gamma_{kl}\frac{\partial x^k}{\partial x'^i}\frac{\partial x^l}{\partial x'^j}.
\end{align}
The objects that transform covariantly under this are then the extrinsic curvature $K_{ij}$, the spatial curvature of the foliation leafs $R_{ijkl}$, and the acceleration vector $a_i=\partial_i \log N$, where\footnote{Here $R_{ijkl}$ and $\nabla_i$ are covariant objects constructed out of the spatial metric $\gamma_{ij}$.}
\begin{align}
    K_{ij}=\frac{1}{2N}\left(\partial_t \gamma_{ij}-\nabla_i N_j-\nabla_j N_i\right).
\end{align}

The action of \Horava Gravity is then written as
\begin{align}\label{eq:action_HG}
    S_{\rm HG}=\frac{1}{16\pi G}\int dt d^Dx\ N \sqrt{\gamma}\left[K_{ij}K^{ij}-\lambda K^2 -{\cal V} \right]
\end{align}
where $\sqrt{|g|}=N\sqrt{\gamma}$, $D=d-1$ is the dimension of the spatial leafs, $\lambda$ is a dimensionless coupling, and ${\cal V}$ contains only spatial derivatives. Since $K_{ij}$ is the only covariant tensor involving time derivatives, the action is explicitly second order in them and avoids the presence of Orstrogradsky ghosts for all observers, since no coordinate transformation of the form \eqref{eq:fdiff} can tilt the time direction. Note that now $[G]_S=z-D$. In the particular case in which ${\cal V}=R-2\Lambda$ and $\lambda=1$, this action reproduces the $D+1$ decomposition of the Einstein-Hilbert action \cite{Gourgoulhon:2007ue}, with the symmetry group enhanced to full diffeomorphisms. One thus defines this as the naive GR limit of the theory.

In Ho\v rava gravity instead, the potential ${\cal V}$ contains all possible terms up to scaling dimension $2z$ constructed out of $R_{ijkl}$, $\nabla_i$ and $a_i$, such as $R, a_i a^i, R^2, R_{ij}R^{ij}, a_i a_j R^{ij}, \dots$ so that the propagator inherits the behavior in \eqref{eq:prop_lifshitz}. A simple computation using the scaling \eqref{eq:lifshitz_scaling} then allows to compute the superficial degree of divergence of a generic Feynman diagram in the UV to be
\begin{align}\label{eq:degree_divergence_horava}
    {\rm div}=2z+2(D-z)L=2z-2L[G]_s.
\end{align}
This result mimics that of Quadratic gravity in some manner. For $z=D$ we have $[G]_s=0$ and the theory becomes power-counting renormalizable, with all divergences corresponding to operators with scaling dimension $[{\cal O}]_s\leq 2z$. These correspond exactly to two time derivatives and up to $2z$ spatial derivatives, per \eqref{eq:lifshitz_scaling}. For $z>D$ we find a super-renormalizable theory. We will take $z=D$ hereinafter, so that the potential contains up to $2D$ spatial derivatives. This corresponds in $3+1$ dimensions to $z=3$ and terms with up to six derivatives.

Note that in general, and in particular for $D=z=3$, the potential ${\cal V}$ contains a large number of terms. Due to this, \Horava originally proposed a condition of detailed balance to reduce its complexity. It amounts to the potential being obtained as the variation of a higher dimensional super-potential, and reduces the number of terms in ${\cal V}$ to six. However, it also brings down a term which violates parity, and a non-vanishing cosmological constant of the wrong sign. Due to this, we will not consider detailed balance here. Instead, we will study two flavors of \Horava Gravity, the projectable and non-projectable theories, distinguished by the way in which they deal with the time reparametrization invariance within \eqref{eq:fdiff}. For more details about the problems found in the detailed balanced theory, see \cite{Sotiriou:2010wn}.

\subsection{Projectable Ho\v rava Gravity}\label{sec:pHG}

In order to reduce the number of terms in the potential, \Horava also proposed the lapse to depend only on time\footnote{The term projectable seems to come from the fact that under this condition, $x^\mu$ becomes a projectable vector field satisfying $\partial^\mu N \frac{\partial}{\partial x^\mu}=N \frac{\partial}{\partial q}$, where $q\in \mathbb{R}$ numbers the foliation leaf and satisfies $\frac{\partial t}{\partial q}=\log(N)$. Note that $q(t)$ is bijective provided that $N$ is monotonous. This is not true in general if $N=N(t,x)$.} $N=N(t)$. This not only enforces the vanishing of any term involving $a_i$, so that the potential is a function of $R_{ijkl}$ and its derivatives only, but also allows to fix $N=1$ as a gauge choice for the time reparametrization invariance in \eqref{eq:fdiff}. The remaining theory enjoys a simplified gauge symmetry consisting on time-dependent spatial diffeomorphisms only
\begin{align}\label{eq:spatial_diff}
    x^i\rightarrow x'^i(t,x).
\end{align}

The action of projectable Ho\v rava Gravity (pHG) thus simplifies to
\begin{align}\label{eq:action_pHG}
    S_{\rm pHG}=\frac{1}{16\pi G}\int dt d^Dx\ \sqrt{\gamma}\left[K_{ij}K^{ij}-\lambda K^2 - {\cal V}_p   \right],
\end{align}
where ${\cal V}_p$ is strongly simplified from the general case. In $D=2$ and $D=3$ it reads
\begin{align}
  \label{eq:V_phg_2}  D=2, \quad {\cal V}_p&= 2\hat \Lambda + \mu R^2\\
    D=3, \quad {\cal V}_p&=2\hat \Lambda-\eta R +\mu_1 R^2 + \mu_2 R_{ij}R^{ij}\nonumber \\
   \label{eq:V_phg_3} &+\n_1 R^3 +\n_2 R R_{ij}R^{ij} + \n_3 R^i_j R^j_kR^k_i+\n_4 \nabla_i R \nabla^i R +\nu_5 \nabla_i R_{jk} \nabla^i R_{jk},
\end{align}
where we have used Bianchi identities to get rid of all other possible equivalent terms. Note that in two (three) dimensions, the Riemann tensor is completely determined by the scalar curvature (and the Ricci tensor). We have also used the Gauss-Bonnet theorem to get rid of the term $R$ in $D=2$\footnote{For even $D$, the spatial integral of the spatial Gauss-Bonnet density ${\cal G}$ is a constant independent of the local values of the fields. Hence the quantity
\begin{align}
    \int dt d^Dx\sqrt{\gamma} \ {\cal G}\propto \int dt {\cal X}, 
\end{align}
where ${\cal X}$ is the Euler's characteristic of ${\cal M}$, is also independent of the fields and does not contribute to local dynamics. Note that this is only true because the lapse has been fixed to a constant value, and thus it is a property of the projectable model \emph{only}. In a general case, $N(t,x)$ is space dependent and cannot be factored out of the spatial integral.}. For concreteness we focus now on the realistic case of $D=3$, but we will come back to the $D=2$ case later when discussing quantization.

Due to the reduced symmetry with respect to GR, the action \eqref{eq:action_pHG} -- although it is also true for \eqref{eq:action_HG} in general -- propagates an extra scalar (s) degree of freedom, on top of the usual transverse-traceless (tt) graviton. Expanding around flat space-time -- corresponding to $N^i=0, \gamma_{ij}=\delta_{ij}$, and $\hat \Lambda=0$ -- both modes have a positive-definite kinetic term in the UV, and hence they are not ghosts, as long as $G>0$ and $\lambda\in (-\infty, \frac{1}{3})\cup(1,\infty)$. Their dispersion relations read
\begin{align}
    &\omega_{tt}^2=\eta k^2 + \mu_2 k^4 + \nu_5 k^6,\\
    &\omega_s^2=\frac{1-\lambda}{1-3\lambda}\left(-\eta k^2 + (8\mu_1+3\mu_2)k^4+(8\n_4 + 3 \n_5)k^6 \right).
\end{align}
These are regular in the UV under the previously mentioned conditions, but lead to pathologies in the IR, where they are dominated by the terms proportional to $k^2$, which have opposite sign. Imposing $\eta>0$, so that the graviton dispersion relation is stable, leads to a tachyonic behavior at low energies for the scalar mode. This implies that flat space is not a stable vacuum of the theory. Attempting to suppress this instability by tuning the values of $\lambda$ or $\eta$ leads to problems. Choosing $\lambda\sim 1$ implies exiting the domain where perturbation theory is under control, while setting $\eta=0$ -- or expanding around a curved background -- produces a theory which departs from GR at low energies \cite{Blas:2010hb}.

In higher dimensions, the situation is equivalent to the one described here. The dispersion relations are regular in the UV, but imposing a healthy behavior for the tt mode leads to either instabilities or strong coupling in the scalar sector. In $D=2$ the tt mode is absent, but the problem persist. A propagating scalar degree of freedom remains, with dispersion relation
\begin{align}\label{eq:disp_scalar_d2}
    \omega_s^2=\frac{4\mu (1-\lambda) k^4}{1-2\lambda},
\end{align}
which has no contribution quadratic in $k$, due to the fact that the term proportional to $R$ in the action is a total derivative. This leads to similar conclusions about instability and strong coupling as discussed above.

In spite of its problems at low energies, pHG remains interesting due to its fundamental properties. Although simpler than the full non-projectable model, it retains most of its particularities. Moreover, pHG is renormalizable beyond power-counting in any space-time dimension \cite{Barvinsky:2015kil}, and a explicit computation of its renormalization group flow in the absence of matter unveils asymptotically free fixed points in $D=2$ \cite{Barvinsky:2017kob} and $D=3$ \cite{Barvinsky:2021ubv}. Some authors have also suggested that the possibility of entering strong coupling at low energies, although a disaster for perturbative computations, can be seen as an opportunity for dynamical emergence of Lorentz invariance at low energies through a phase transition á la QCD, where the scalar mode condenses and leaves the spectrum \cite{Blas:2018flu}. It has also been argued that it might take the role of dark matter \cite{Mukohyama:2009mz}. However, a proper assessment of this behavior would require the use of lattice or other non-perturbative methods.

\subsection{Non-projectable Ho\v rava Gravity}\label{sec:npHG}

In spite of its simpler potential, imposing an extra symmetry just to simplify the theory down to pHG is an adhoc ingredient. If projectability is abandoned, so that $N=N(t,x)$, the number of allowed terms in ${\cal V}$ grows quickly with $D$, and all of them must be included by renormalizability arguments \cite{Blas:2009qj}. However, this is the \emph{true} formulation of \Horava Gravity -- a power-counting renormalizable theory of gravitation with Lifshitz scaling \eqref{eq:lifshitz_scaling} in the UV and no extra constraints. Moreover, the addition of the new terms involving $a_i$ solve the problem of the instability of the scalar mode at low energies found in pHG \cite{Blas:2009yd}. Although they will also lead to explicit Lorentz violations in the IR, these are compatible with current observations and experimental constraints by cosmological and astrophysical results, as we will see in section \ref{sec:constraints}.

The action of non-projectable \Horava Gravity can be written in the following form
\begin{align}\label{eq:np_action}
    S_{\rm npHG}=\frac{1}{16\pi G}\int dt d^Dx\ N \sqrt{\gamma}\left[K_{ij}K^{ij}-\lambda K^2+\eta R +\alpha a_i a^i-2\hat \Lambda -\sum^{D}_{2} \frac{L_{2n}}{M_*^{2n-2}} \right],
\end{align}
where $M_*$ is the scale at which higher derivative terms start dominating the dynamics, and $L_{2n}$ contains all possible independent terms of dimension $2n$ constructed by using $R_{ijkl}$, $\nabla_{i}$, and $a_i$ -- operators with an odd number of derivatives are forbidden by parity. Their number grows quickly with $D$, but we can look at the full expression in $D=2$ in order to compare with the simplicity of the projectable case in \eqref{eq:V_phg_2}. In this case, we only have $L_4$, since $z=D=2$, which reads
\begin{align}
    L_4=\mu R^2+\rho_1 \Delta R \rho_2 R a_i a^i + \rho_3 (a_i a^i)^2+\rho_4 a_ i a^i \nabla_j a^j +\rho_5 (\nabla_j a^j)^2 + \rho_6 \nabla_i a_j \nabla^i a^j.
\end{align}
We see that while pHG only had two terms in its potential, here we find -- counting also those written explicitly in \eqref{eq:np_action} -- ten nonequivalent operators, which represent an increase of roughly an order of magnitude. This is even more dramatic in $D=3$, and certainly complicates the performance of explicit analytical computations in the theory.

As we mentioned before, the addition of the new terms depending on $a_i$ can cure the instability of the scalar mode. This is clearly seen in $D=3$. Fixing the cosmological constant to vanish and expanding around flat space-time, the dispersion relations read now 
\begin{align}
    &\omega_{tt}^2=k^2 +{\cal O}(k^4),\\
    &\omega_s^2=\frac{1-\lambda}{1-3\lambda}\left(\frac{2}{\alpha}-1\right) k^2+{\cal O}(k^4),
\end{align}
where we have chosen\footnote{This is always possible in the absence of matter by a rescaling of the spatial coordinate.} $\eta=1$. The ${\cal O}(k^4)$ terms are controlled by the operators in $L_4$ and $L_6$. Their exact form is irrelevant for our present discussion but one should keep in mind that demanding a regular propagator at high energies will impose constraints in the values of the couplings accompanying them. Note that at low energies, the dispersion relation of the tt mode remains the same as in the projectable case. This is trivial to confirm, since the only new term at the two-derivative level in \eqref{eq:np_action} is $a_ia^i$, which contributes only to the scalar sector at quadratic order. The dispersion relation of the scalar mode does instead get modified, picking a new term depending on $\alpha$. For values $0<\alpha<2$ -- and as long as $\lambda\in (-\infty, \frac{1}{3})\cup(1,\infty)$ -- the scalar mode is also regular at low energies, thus avoiding the instability, and in particular allowing for the possibility of maintaining perturbative control at all energies.

Note however that due to this term, the propagation speed of the scalar mode will in general differ from that of the tt mode
\begin{align}\label{eq:propagation_speed}
    \frac{c^2_s}{c_{tt}^2}=\frac{1-\lambda}{1-3\lambda}\left(\frac{2}{\alpha}-1\right),
\end{align}
which means that Lorentz invariance is generically broken at all energies. This leads to a very rich phenomenology, different from the GR one, that can be constrained by observations and experiments in the gravity sector (cf. section \ref{sec:constraints}).

Finally, let us add some words about the value of the high-energy scale $M_*$, following the discussion in \cite{Sotiriou:2010wn}. Note that in order for higher derivatives to effectively make the theory perturbatively renormalizable at all scales, $M_*$ has to be smaller than the corresponding energy scale at which the low energy theory -- given by $M_* \rightarrow \infty$ -- enters strong coupling. Similarly to the analysis for GR, this happens when the scale accompanying cubic and higher point interactions becomes larger than one. Expanding \eqref{eq:np_action} around flat space and canonically normalizing the kinetic terms, we can see that there are several scales involved in this interplay, so that we obtain $M_*\ll {\rm Min}\{\sqrt{\eta} M_P,\sqrt{|\lambda-1|}M_P\}$. Taking into account Solar system tests on $\lambda$ -- again, see more in section \ref{sec:constraints} --, gives $M_*< 10^{15}{\rm GeV}$. Assuming that this same scale percolates to any matter action coupled to Ho\v rava gravity allows to improve this constraint to a double side bound. In particular, the absence of time delay in $\gamma$-rays of different frequency emmited by distant astrophysical sources \cite{MAGIC:2007etg,FermiGBMLAT:2009nfe} implies $M_*>10^{11}{\rm GeV}$ and hence
\begin{align}\label{eq:value_Mstar}
    10^{11} {\rm GeV}< M_* < 10^{15} {\rm GeV}.
\end{align}
The existence of such a window of energies can be thought as one of the successes of the proposal of Ho\v rava gravity. While small, it allows for the presence of Lorentz violations in the gravitational sector without distorting well-established experimental and observational phenomena.


\section{Ho\v rava Gravity in the ultra-violet}

The original goal of \Horava Gravity was to provide a UV completion of GR -- a formulation of a gravitational theory which solves the main problem of Einstein's theory, its non-renormalizable character. Indeed, as we have seen, this is easily achieved naively by extending the action with higher derivative terms, which lead to power-counting renormalizability. However, that is not enough to conclude that \Horava gravity is a theory of Quantum Gravity, since it is a gauge theory. A successful quantization scheme not only has to provide terms with the right scaling dimension, but also has to preserve the symmetries of the action -- the FDiff transformations \eqref{eq:fdiff} in this case --, and even more importantly, all divergences must correspond to local operators. Even if this is satisfied, a discussion on the renormalization group flow of the theory is necessary in order to understand the behavior of the degrees of freedom at all energies, and to assess whether the proposal can be realized within perturbative control. As we will see in what follows, all these are subtle issues in the case of \Horava gravity, starting with the algebra of constraints, which include second-class constraints that obstruct a straightforward quantization of the non-projectable model.

\subsection{The algebra of constraints}\label{sec:hamiltonian}

In order to obtain the constraint structure of the theory, we turn to the Hamiltonian formalism by introducing the canonical pairs $(\gamma_{ij},\pi_{ij})$, $(N_i,p_i)$, $(N,p_0)$. The fields $N$ and $N_i$ appear in the action without time derivatives, and hence we have the following primary constraints for their conjugate momenta \cite{Donnelly:2011df}
\begin{align}
    p_0=0,\quad p_i=0.
\end{align}
On the other hand, the momentum conjugated to the spatial metric is
\begin{align}
    \pi_{ij}=\sqrt{\gamma}\left(K_{ij}-\lambda K \gamma_{ij}\right).
\end{align}

This allows to write the Hamiltonian of \Horava gravity as
\begin{align}\label{eq:Hamiltonian}
    H=\int_{\Sigma} d^dx \left( N {\cal H}_0 +N^i{\cal H}_i + v^ip_i + v p_0 \right),
\end{align}
where we have assumed that the spatial manifold $\Sigma$ has no boundary, and introduced Lagrange multipliers $v$ and $v_i$ to impose the primary constraints. Also
\begin{align}
    &{\cal H}_0=\frac{1}{\sqrt{\gamma}}\left(\pi_{ij}\pi^{ij}+\frac{\lambda}{1-3\lambda}\pi^2 + \gamma {\cal V}\right),\\
    &{\cal H}_i=-2 \gamma_{ik}\nabla_j \pi^{jk}.
\end{align}

Let us focus first on the primary constraint $p_i=0$. Its preservation in time implies a secondary constraint
\begin{align}
    C_i=\frac{\delta H}{\delta N^i}={\cal H}_i=0.
\end{align}
This is identical to the usual momentum constraint of GR. It generates spatial diffeomorphisms of $\gamma_{ij}$ and $\pi^{ij}$, and its Poisson bracket algebra closes on-shell when the constraint itself is satisfied at equal times
\begin{align}\label{eq:Ci_bracket}
    \{C_i(x),C_j(y)\}=C_i(y)\partial_j\delta(x,y)+C_j(x) \partial_i\delta(x,y).
\end{align}
Its preservation in time is not obvious, but it can be shown by defining the smeared constraint
\begin{align}
    \tilde{C}_i=C_i+p_0 \nabla_i N  +{\cal L}_{N^i} p_i,
\end{align}
where ${\cal L}_{N^i}$ is the Lie derivative along $N^i$. Note that the extra terms are proportional to the primary constraints and thus they vanish on-shell. This new expression now generates spatial diffeomorphisms on all metric variables $\gamma_{ij}$, $N$, and $N_i$, and hence its Poisson bracket with the Hamiltonian, which is FDiff invariant, vanishes. Preservation in time of $\tilde C_i$ is then automatic, from which preservation of $C_i$ follows. We conclude that the secondary constraint $C_i=0$ is first-class. The shift vector $N^i$ acts as a Lagrange multiplier in \eqref{eq:Hamiltonian} imposing it and thus it can be fixed to any constant value as a gauge choice. Together with the primary constraint $p_i=0$, this eliminates $2D$ degrees of freedom. 

The role of the second primary constraint $p_0=0$ is more subtle. Its preservation in time also leads to a secondary constraint
\begin{align}\label{eq:c2_constraint}
    C={\cal H}_0-\frac{1}{N}\nabla_i \left(\sqrt{\gamma} N \frac{\delta {\cal V}}{\delta a_i}\right)=0.
\end{align}
In contrast to what happened with the momentum constraint $C_i$, the energy constraint $C$ differs from the GR one. Only if ${\cal V}$ has no $a_i$ dependence, as it exactly happens in GR, we can find a closed Poisson algebra $\{ C(x),C(y)\}=0 $ and $\{C(x),p_0(y)\}=0$ on the mass-shell. However, whenever the second term in \eqref{eq:c2_constraint} is non-vanishing, we find
\begin{align}\label{eq:second_class}
\{C(x),P_0(y)\}\propto \frac{\partial {\cal V}}{\delta a_i \delta a_j},
\end{align}
so that the constraints are second-class. In this case, and although $P_0=0$ still eliminates a degree of freedom, the lapse is not a Lagrange multiplier in \eqref{eq:Hamiltonian} and it cannot be set to a constant value as a gauge fixing. Instead, the secondary constraint \eqref{eq:c2_constraint} can be interpreted as an equation of motion for $N$. Note that on-shell vanishing of $C$ ensures also on-shell vanishing of the Hamiltonian $H$, since the second term in \eqref{eq:c2_constraint} becomes a total derivative when integrated over $\Sigma$. This is a property of theories enjoying time reparametrization invariance.

Counting the number of fields which do not propagate due to the constraints, we therefore see that \Horava Gravity displays
\begin{align}
    \frac{d(d+1)}{2}-2D-1=\frac{(D+1)(D+2)}{2}-2D-1=\frac{D(D-1)}{2}
\end{align}
degrees of freedom. These correspond to the usual transverse-traceless graviton in $D>2$ plus an extra scalar, which appears due to the reduced symmetry of the theory. 

Although the derivation here uses the general form of the action and symmetry found in non-projectable \Horava gravity, its reduction to pHG is automatic. From the beginning we choose $N(t)=1$ by fixing time reparametrization invariance. The derivation above then follows, with the theory containing only the primary and secondary constraints $p_i=0$ and $C_i=0$. The number of propagating degrees of freedom remain the same as in the general case. Note that the presence of the extra scalar implies that, while GR has no local degree of freedom in $2+1$ dimensions -- only a global one fixed by boundary conditions \cite{Witten:1988hc} -- \Horava gravity propagates a local perturbation.

Although quantization of the projectable case can then be done straigthforwardly with standard techniques -- albeit with subtleties related to locality, as we see below --, a similar approach cannot be done for the general non-projectable theory. In that case, two of the constrains are second-class, and their quantization requires the use of Dirac brackets and the formalism developed by Batalin, Fradkin, Fradkina and Vilkovisky for quantization of constrained gauge theories \cite{Fradkin:1975cq,Batalin:1977pb,Fradkin:1977xi}. Even with this formalism at hand, the question on the consistency of a quantum version of non-projectable Ho\v rava gravity is still unanswered.
\subsection{Renormalization of projectable Ho\v rava Gravity}\label{sec:renorm_pHG}

As we have seen before, in the case of pHG time reparametrization invariance is gauged away of the theory by fixing the value of the lapse to a constant $N(t)=1$ that is chosen to unity for convenience. In this case, the second-class constraints \eqref{eq:second_class} inherited from the vanishing of the momentum conjugate to $N$ are absent. The only remaining constraint is first-class, and quantization can be attempted in the standard way, by introducing a gauge fixing condition in the path integral in Lagrangian formulation. However, things are not straightforward, since most gauge choices lead to pathologies in the structure of divergences already at the one-loop level \cite{Barvinsky:2015kil,Barvinsky:2023mrv}. In order to see this, we first expand the remaining variables around flat space-time\footnote{Note that here we are not normalizing the kinetic term of the graviton.}
\begin{align}\label{eq:flat_space_perturbation}
    \gamma_{ij}=\delta_{ij}+h_{ij},\quad N^i=0+n^i,    
\end{align}
and retain terms in the action up to fourth order in perturbations, which are enough for one-loop computations. The problem raised by the symmetry structure can be seen quickly in $D=2$ by choosing the straightforward gauge condition $N^i=0$. In this gauge, the propagator for gravitational perturbations picks up a contribution
\begin{align}
    \langle h_{ij}(\omega,k)h_{kl}(-\omega,-k)\rangle  \supset \frac{1}{\omega^2}.
\end{align}
This term is dumped in frequency, but not in momentum, so its coordinate representation after Fourier transform
\begin{align}\label{eq:non_local_div}
    \langle h_{ij}(t_1,x_1)h_{kl}(t_2,x_2)\rangle\supset |t_1-t_2|\delta^{(2)}(x_1-x_2),
\end{align}
is singular at all times, in sharp contrast to the localized singularities provided by local QFT's. The origin of this issue can be traced back to the role of the perturbation of the shift in the action, which induces non-diagonal couplings already at the two-field level of the form 
\begin{align}\label{eq:hN_terms}
    \partial_t n_i \partial_j h^{ij},
\end{align}
and similar. These terms contain odd numbers of time derivatives, which give raise to nested structures in the propagator of the schematic form $\left[\omega^2 (\omega^2- c^2 k^4)\right]^{-1}$ for some $c$. 

These odd-derivative terms are however unavoidable, since they are induced by the form of the FDiff preserving covariant time derivative, which is encoded within $K_{ij}$ and becomes explicit when expanding in perturbations. On a generic tensor $T$ it acts as
\begin{align}
    D_t T=\partial_t T - {\cal L}_{N^i}T.
\end{align}
Due to the anisotropic scaling \eqref{eq:lifshitz_scaling}, the shift vector and its perturbation acquire a non-trivial scaling dimension $[N^i]=[n^i]=D-1$ and hence the object $\partial_i n_j$ scales as a time derivative.

A possible way out of the issue is to introduce terms in the gauge fixing choice that cancel the terms of the form \eqref{eq:hN_terms}. Since the gauge fixing function is a vector in this case, this implies that it must take a form $F_i=a_g\partial_t n_i+b_g \partial_j h_i^{j}$, where $a_g$ and $b_g$ are arbitrary coefficients. This is actually what we find if we open up the usual DeDonder gauge fixing used in GR \eqref{eq:dedonder} in ADM variables. However, in this case this term is not homogeneous in scaling dimension, due to the anisotropic scaling \eqref{eq:lifshitz_scaling}. Writing a homogeneous term requires acting on $h_{ij}$ with more derivatives in order to compensate the scaling of the time derivative in the first term. This, however, cannot cancel the problematic pieces and, more importantly, will exceed the critical scaling dimension $2D$ of the bare Lagrangian when squared to add it to the path integral. The next possible approach would be to construct the gauge breaking term in the path integral through the action of a local operator in the form
\begin{align}
    S_{gf}=\int dt d^Dx\ F^i{\cal O}_{ij} F^j.
\end{align}
However, the index structure fixes ${\cal O}_{ij}=(-\Delta)^{a}(-\delta_{ij}\Delta-\xi \partial_i \partial_j)$, with $\xi\neq -1$, $\Delta=\delta^{ij}\partial_i\partial_j$ and $a$ chosen to get a total scaling dimension of $2D$. The same problem arises again -- it is impossible to write a homogeneous term including $\partial_t n^i$ without exceeding the total scaling dimension $2z$, \emph{unless we choose $a< 0$}.

Precisely this last observation is the resolution of the conundrum. If we allow the operator ${\cal O}_{ij}$ to have negative dimension, and hence be non-local, a satisfactory gauge-fixing can be constructed, as introduced in \cite{Barvinsky:2015kil}. Successful quantization of pHG can then be achieved in the family of $\sigma\xi$-gauges of the form
\begin{align}
    &F^{i}=\partial_t n^i+\frac{1}{2\sigma}\left({\cal O}^{-1}\right)^{ij}\left(\partial^k h_{kj}-\lambda \partial_j h\right),\\
   \label{eq:Ouperator} &{\cal O}_{ij}=-\left(-\Delta\right)^{2-D}\left(\Delta\delta_{ij}+\xi\partial_i\partial_j\right),
\end{align}
where $\sigma$ and $\xi$ can be chosen at convenience to simplify the resulting propagators, and $\lambda$ is introduced in order to ensure cancellation of all non-diagonal terms involving $h_{ij}$ and $n^i$.

Let us show briefly how this does the trick in $D=2$ \cite{Barvinsky:2017kob}. For a longer discussion involving the choice $D=3$ see \cite{Barvinsky:2015kil,Barvinsky:2019rwn,Barvinsky:2021ubv}. In this case, the propagators read
\begin{align}
    &\langle n^i(\omega , k) n^j(-\omega,-k)\rangle= \frac{\kappa}{\sigma}\left(k^2\delta_{ij}-k_i k_j\right){\cal P}_1(\omega,k)+\frac{\kappa (1+\xi)}{\sigma}k_ik_j {\cal P}_2(\omega,k),\\
    \nonumber &\langle h_{ij}(\omega ,k)h_{kl}(-\omega,-k)\rangle=4 \kappa \delta_{ij}\delta_{kl}\left[\frac{1-\lambda}{1-2\lambda}{\cal P}_s (\omega,k)-{\cal P}_1(\omega,k)\right]+2 \kappa (\delta_{ik}\delta_{jl}+\delta_{il}\delta_{jk}){\cal P}_1(\omega,k)\\
    &+ \frac{4 \kappa}{k^2} (\delta_{ij}k_k k_l+\delta_{kl}k_ik_j)\left[{\cal P}_1(\omega,k)-{\cal P}_s(\omega,k)\right]+\frac{4\kappa}{k^4}k_ik_jk_kk_l\left[\frac{1-2\lambda}{1-\lambda}{\cal P}_s(\omega,k)-2{\cal P}_1(\omega,k)+\frac{{\cal P}_2(\omega,k)}{1-\lambda}\right],
\end{align}
where $\kappa=8\pi G$ is the effective gravitational coupling. The pole structures ${\cal P}_i(\omega,k)$ are given by
\begin{align}
    &{\cal P}^{-1}_s(\omega,k)=\omega^2-\frac{4\mu (1-\lambda) k^4}{1-2\lambda},\\
    &{\cal P}^{-1}_1(\omega,k)=\omega^2-\frac{k^4}{2\sigma}\\
    &{\cal P}^{-1}_2(\omega,k)=\omega^2-\frac{(1-\lambda)(1+\xi)k^4}{\sigma}.
\end{align}
The first one corresponds to the physical pole of the propagating scalar degree of freedom, with dispersion relation \eqref{eq:disp_scalar_d2}, while the others represent gauge degrees of freedom. They are compensated by the ones in the ghost sector of the theory. This is obtained in the standard way, by introducing a pair of anti-commuting fields $c^i$ and $\bar{c}_j$, whose action is constructed by acting on $F^i$ with a gauge transformation, replacing the gauge parameter by the field $c^i$. Their propagator reads  
\begin{align}
    \langle c^i(\omega, k) \bar{c}_j(-\omega ,-k)\rangle= \kappa \delta^i_{j}{\cal P}_1(\omega,k)+\kappa k^ik_j \left[{\cal P}_2(\omega,k)-{\cal P}_1(\omega,k)\right],
\end{align}
and indeed contains only the gauge poles. A more careful discussion, based on the use of BRST invariance, can be found in \cite{Barvinsky:2015kil}. Once all the propagators display regular poles, the superficial degree of divergence follows the formula \eqref{eq:degree_divergence_horava} \cite{Anselmi:2007ri}. Although here we have used an expansion around flat spacetime for simplicity of the discussion, the same properties remain true around arbitrary backgrounds, provided that the background field method is used \cite{Barvinsky:2017zlx}. This ensures that all divergences are local -- thanks to regularity of propagators -- and proportional to gauge invariant operators already in the bare action. The theory is thus fully renormalizable. This indicates that if attempted, a full quantum computation in any gauge would have produced a pure local answer, with all possible non-local divergences, such as the one in \eqref{eq:non_local_div}, cancelling in the final result thanks to gauge invariance. Of course, it was impossible to know this a priori.

As a final note, one could be worried about the presence of a non-local operator in the gauge fixing term, since the initial theory is perfectly local. However, its presence is actually harmless. This can be seen by noting that the non-local operator actually cancels everywhere in the gauge fixing action but in a single term, the kinetic term of the shift
\begin{align}
    \sim \int dt d^Dx\ \partial_t n^i {\cal O}_{ij} \partial_t n^j,
\end{align}
which can be cast in a local form by introducing an auxiliary field $p_i$
\begin{align}
    \int dt d^Dx\left[ \frac{1}{2\sigma} p_i \left({\cal O}^{-1}\right)^{ij}p_j - i p_i\partial_t n^i \right].
\end{align}
Upon integration of $p_i$, which enters as a momentum variable conjugated to $n^i$ -- thus its name --, the original gauge-fixing term is recovered. However, in this form the action is explicitly local, and all the propagators remain regular, including the ones involving $p_i$. The fact that $p_i$ is a momentum variable can be seen in retrospective as a hint on how to approach the most complicated task of quantizing the non-projectable theory, but we will come back to this later.

\subsubsection{Asymptotic Freedom in projectable \Horava Gravity}

Beyond renormalizability, the big question mark in the consistency of quantum \Horava gravity is the understanding of its UV structure. Previous works obtained partial results in a simplified model obtained by a conformal reduction of the metric \cite{Benedetti:2013pya}, and through the calculation of matter loops to the renormalization group (RG) flow of the gravitational couplings \cite{DOdorico:2014tyh}. However, it was not until the proof of renormalization in \cite{Barvinsky:2015kil} that a explicit gravitational computation in the full projectable theory was performed. The calculation is however very lengthy due to the large number of terms proliferating in intermediate steps. This can be understood by noting that a generic vertex with $N_h$ gravitational legs and $N_n$ shift legs has $2N_h+N_n$ free indices. Although momentum conservation typically alleviates the problem a lot, one can estimate the maximum number of independent terms in a given vertex to be then $\sim (2N_h+N_n)!$, which quickly gets out of hand.

\begin{figure}
	\centering
	\includegraphics[width=0.5\textwidth]{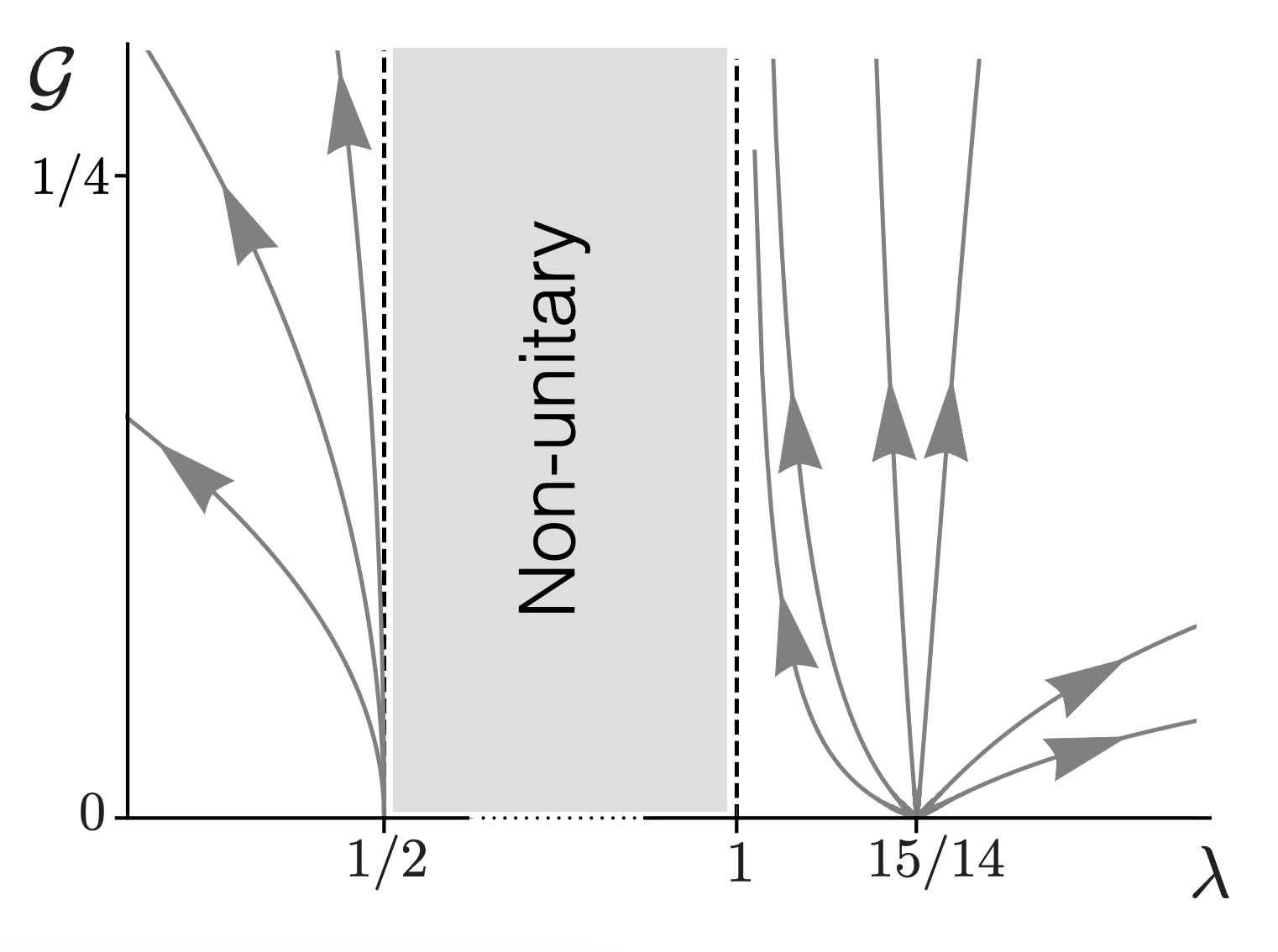}
	\caption{Renormalization group flow of the essential couplings of pHG in $D=2$. The non-unitary region corresponds to values of $\lambda$ for which the pole in the propagator of the scalar degree of freedom is in the wrong part of the complex plane. The arrows point towards the IR. Figure from \cite{Barvinsky:2017kob}.}
        \label{fig:rgflow_d2}
\end{figure}

Nonetheless, a first computation in $D=2$ was performed in \cite{Griffin:2017wvh}, where the running of the cosmological constant was obtained. Later, the full RG flow of the essential couplings of the theory was computed in \cite{Barvinsky:2017kob} using standard Feynman diagrams. Essential couplings are those accompanying on-shell operators, whose value is invariant under the addition of the equations of motion. In the case of pHG, they are obtained by noting that the equations of motion of the theory imply the vanishing of a quantity which actually corresponds to the on-shell Hamiltonian
\begin{align}
  H=\int dt d^Dx\sqrt{\gamma}\left[K_{ij}K^{ij}-\lambda K^2 + {\cal V}\right]=0.
\end{align}
Imposing that the quantum effective action $\Gamma$ is invariant under $\Gamma\rightarrow \Gamma+\epsilon H$ for arbitrary constant $\epsilon$ singles out the essential couplings. In $D=2$ these are $\lambda$ and ${\cal G}_2=8\pi G/\sqrt{\mu}$, where the coefficient $8\pi$ is added for convenience. This latter coupling controls the interaction strength and thus it is the one which must remain small to retain perturbative control of the theory. Their $\beta$-functions\footnote{We define the $\beta$-function of a coupling $g$ as
\begin{align}
    \beta_g=k_* \frac{dg}{dk_*},
\end{align}
where $k_*$ is the reference momentum scale characterizing the running.} are gauge invariant and read
\begin{align}
    &\beta^{D=2}_\lambda = \frac{15-14\lambda}{64\pi}\sqrt{\frac{1-2\lambda}{1-\lambda}}\ {\cal G}_2,\\
    &\beta_{{\cal G}_2}= -\frac{(16-33\lambda+18\lambda^2)}{64\pi (1-\lambda)^2}\sqrt{\frac{1-\lambda}{1-2\lambda}}\ {\cal G}_2^2.
\end{align}
The corresponding flow is shown in figure \ref{fig:rgflow_d2}. Note that the $\beta$-functions diverge/cancel in the limits of the unitarity region $\lambda=1$ and $\lambda=\frac{1}{2}$, where the symmetry of the kinetic term is enlarged. The former corresponds to the restoration of diffeomorphisms, while in the latter an anisotropic local scale invariance \cite{Arav:2014goa}, given by \eqref{eq:lifshitz_scaling} with a parameter $\chi(t,x)$ depending on the space-time point, emerges in the kinetic term. We see that for values $\lambda>1$, the RG flow displays a UV attractive and asymptotically free fixed point at $\lambda=15/14$, where the gravitational coupling ${\cal G}_2$, which controls interaction with matter, vanishes. The presence of this point is compatible with trajectories that flow towards $\lambda=1$ in the IR, which indicates a possible restoration of Lorentz invariance at low energies. However, in this limit we also have ${\cal G}_2\rightarrow \infty$, thus losing perturbative control and entering into a strongly coupled regime, a behavior similar to that of QCD. On the other side of the non-unitary region, the surface $\lambda=1/2$ also seems to correspond to a fixed point of the RG flow. However, as suggested in \cite{Barvinsky:2017kob}, the true expansion parameter controlling correlators seems to be instead ${\cal G}_2/\sqrt{1-2\lambda}$, whose running freezes at a constant value when $\lambda \rightarrow 1/2$, so that nothing can be concluded in the present state-of-the-art of the calculation. Assessing its behavior would require to include contributions from higher loops or matter interactions.

In $D=3$ things are somewhat different. Although a diagramatic approach is possible, its complexity grows quickly. Nevertheless, the $\beta$-functions for $G$ and $\lambda$ were obtained with this method in \cite{Barvinsky:2019rwn}. A full computation of the remaining running functions was performed in \cite{Barvinsky:2021ubv}, using instead an extension of the Schwinger-Dewitt \cite{Barvinsky:1985an,DeWitt:2003pm}, or Gilkey-Seeley \cite{Gilkey:1995mj,Gilkey:1975iq}, heat kernel method around an arbitrary background space-time. This provides an effective resummation of the perturbative series, allowing to obtain the UV divergences accompanying full non-linear counterterms \cite{Vassilevich:2003xt}, in contrast to the Feynman diagrammatic technique, which perturbs close to flat space. A detailed review on this approach can also be found in \cite{Barvinsky:2023mrv}.

Out of the couplings in the action, with potential \eqref{eq:V_phg_3}, the authors of \cite{Barvinsky:2021ubv} identify six independent essential couplings in $D=3$, which can be chosen to be
\begin{align}
    {\cal G}_3=\frac{8\pi G}{\sqrt{\nu_5}},\quad \lambda, \quad u_s=\sqrt{\frac{\nu_s}{\nu_5}},\quad v_a=\frac{\nu_a}{\nu_5}, \quad a=1,2,3,
\end{align}
where $\nu_s=(8\nu_4+3\nu_5)(1-\lambda)/(1-3\lambda)$. The $\beta$-function of $\lambda$ was computed in \cite{Barvinsky:2019rwn} and confirmed in \cite{Barvinsky:2021ubv}, and results to be surprisingly simple
\begin{align}
    \beta^{D=3}_\lambda=\frac{{\cal G}_3}{120\pi^2 (1-\lambda)(1+u_s)u_s}\left[27(1-\lambda)^2+3u_s (11-3\lambda)(1-\lambda)-2u_s^2(1-3\lambda)^2\right],
\end{align}
in contrast to the complexity of those of the rest of essential couplings. Those are also polynomials in $u_s$, but the coefficient of every term is a complicated function of the rest of couplings
\begin{align}
    &\beta_{{\cal G}_3}=\frac{{\cal G}_3^2}{26880\pi^2(1-\lambda)^2 (1-3\lambda)^2 (1+u_s)^3 u_s^3}\sum_{n=0}^7 u_s^n P^{{\cal G}_3}_n(\lambda,v_1,v_2,v_3),\\
    &\beta_{a}=\frac{A_a{\cal G}_3}{26880\pi^2(1-\lambda)^3 (1-3\lambda)^3 (1+u_s)^3 u_s^5}\sum_{n=0}^7 u_s^n P^{a}_n(\lambda,v_1,v_2,v_3).
\end{align}
Here the index $a$ labels the rest of the couplings $a=(u_s,v_1,v_2,v_3)$, the explicit form of the polinomials $P^{{\cal G}_3}_n(\lambda,v_1,v_2,v_3)$ and $P^{a}_n(\lambda,v_1,v_2,v_3)$ is given in \cite{Barvinsky:2021ubv}, and the prefactor $A_a$ reads in each case
\begin{align}
    A_{u_s}=u_s(1-\lambda), \quad A_{\nu_1}=1,\quad A_{\nu_2}=A_{\nu_3}=2.
\end{align}
Notice that in similar occurrence to $D=2$, the gravitational coupling ${\cal G}_3$ factorizes and enters into the $\beta$-functions as an overall factor. Note as well that all $\beta$-functions are singular when $\lambda=1/3$, $\lambda=1$ or $u_s=0$. The first two values mimic the behavior in $D=2$, corresponding to the limits of the unitary region and an enlargement of the underlying symmetry. The last option leads to a degenerate kinetic term for the scalar field, with the $k^6$ term in the dispersion relation vanishing.

The structure of fixed points of the RG flow has not been deeply studied in this case, due to the complicated expressions of the polinomial factors $P^{{\cal G}_3}_n(\lambda,v_1,v_2,v_3)$ and $P^{a}_n(\lambda,v_1,v_2,v_3)$. However, the work of \cite{Barvinsky:2021ubv} shows some preliminary results. In contrast to the behavior of the theory in $D=2$, they report no asymptotically free fixed points at finite $\lambda>1$. Four asymptotically fixed points -- with ${\cal G}_3\rightarrow 0$ -- are found however when $\lambda<1/3$ instead, but all of them are UV repulsive along the $\lambda$-direction.

A different option, proposed by \cite{Gumrukcuoglu:2011xg} early after Ho\v rava's seminal paper, is that the UV fixed point of pHG lays at $\lambda \rightarrow \infty$. This limit, although counter-intuitive, is surprisingly well-behaved and leads to regular equations of motion at the classical level. A large value of $\lambda$ simply enhances the dynamics of the gravitational field against that of matter fields coupled to it. Strikingly, this conjecture seems to be compatible with the $\beta$-functions obtained by \cite{Barvinsky:2021ubv}. For $\lambda\rightarrow \infty$, all running functions are finite, whereas $\beta_\lambda$ becomes proportional to $\lambda$
\begin{align}
    \beta^{D=3}_\lambda=-\frac{3(3-2u_s)}{40\pi^2 u_s}\lambda {\cal G}_3,\quad \lambda \rightarrow \infty,
\end{align}
which indicates a UV attractive (repulsive) behavior for $u_s>3/2$ ($u_s<3/2$). The authors of identify eight fixed points at $\lambda \rightarrow \infty$, three of them being simultaneously UV attractive and asymptotically free. This supports the conjecture of \cite{Gumrukcuoglu:2011xg} and points towards the need of a better understanding of the UV structure of pHG for large values of $\lambda$, a program which was initiated very recently in \cite{Radkovski:2023cew}. There, the full set of tree-level scattering amplitudes is computed in the limit $\lambda\rightarrow \infty$, finding the expected regular behavior that supports the conjecture. Moreover, the authors provide an effective action describing this limit in a regular way, reading
\begin{align} \label{eq:S_inf}
    S_{\lambda\rightarrow \infty}=\int dt d^3x \sqrt{\gamma}\left(K_{ij}K^{ij}-2\zeta K-{\cal V}\right),
\end{align}
where $\zeta$ is a Lagrange multiplier. This action is very similar to the original action of pHG, but imposes the condition $K=0$. It reproduces all the scattering amplitudes at $\lambda\rightarrow \infty$, and requires to take no limit. More importantly, since in the neighborhood of the UV fixed point all remaining couplings remain finite, the action \eqref{eq:S_inf} describes the dynamics of the theory at arbitrary high energies. No running of the couplings nor any extra ingredient needs to be included. The action \eqref{eq:S_inf} is the theory at the fixed point of pHG in $3+1$ dimensions, a true theory of quantum gravity.

\subsection{Towards the renormalization of non-projectable \Horava Gravity}\label{sec:renorm_nphg}

We return now to the question of renormalizability of the non-projectable version of \Horava Gravity. As we have seen, quantization of pHG is carried out in the standard way for gauge theories, by choosing a gauge fixing condition for spatial diffeomorphisms and introducing it in the action together with the Lagrangian for ghosts. Particularities on the construction of the gauge fixing aside, this can be done straigthforwardly because the associated constraints in the Hamiltonian formalism are first-class, so that quantization of pHG does not differ much from that of GR at that level. Attempting to go down the same road with non-projectable \Horava gravity fails down however, due to the presence of second-class constraints -- see section \ref{sec:hamiltonian}. A more complicated quantization scheme -- only recently developed by \cite{Bellorin:2021tkk,Bellorin:2022qeu} -- has to be used, based on the Hamiltonian formalism. Focusing in the particular situation at hand, the way to go is to use the formalism developed by Batalin, Fradkin, and Vilkovisky (BFV) for quantization of theories in the Hamiltonian form \cite{Fradkin:1975cq,Batalin:1977pb}. This was later extended by Fradkin and Fradkina to the case of theories with second class constraints \cite{Fradkin:1977xi}, which fits the case of \Horava Gravity. 

We start by redefining the secondary constraints as 
\begin{align}
    \theta_1=N C=N {\cal H}_0 - \nabla_i \left(\sqrt{\gamma}\ N \frac{\delta {\cal V}}{\delta a_i}\right), \quad \theta_2=p_0=0,
\end{align}
in order to follow the notation of \cite{Bellorin:2022qeu}. Note that the rescaling of $\theta_1$ is harmless, since the lapse $N$ can only vanish point-wise as most. With this, the primary Hamiltonian -- the Hamiltonian \eqref{eq:Hamiltonian} undressed of the constraints -- reads
\begin{align}
    H_0=\int_\Sigma d^dx\ \theta_1.
\end{align}

The BFV formalism for second-class theories then demands the existence of a complete set of functions $G_a$, which must include the first class constraints, satisfying the involutive algebra
\begin{align}\label{eq:involutive}
    \{G_a, G_b\}_D=U^c_{ab} G_c,\quad \{ {\cal H}_0,G_a\}_D = V_a^b G_b,
\end{align}
where $\{,\}_D$ is a Dirac bracket, defined by
\begin{align}
    \{\hat A,\hat B\}_D=\{\hat A,\hat B\}-\{\hat A,\theta_A\}\mathfrak{M}^{-1}_{AB}\{\theta_B,\hat B\}.    
\end{align}
Here the matrix $\mathfrak{M}$ contains the Poisson brackets of the secondary constraints, which for the case of \Horava Gravity takes a triangular form
\begin{align}\label{eq:M_matrix}
    \mathfrak{M}=
    \begin{pmatrix}
    0& \{\theta_1,\theta_2\}\\
    -\{\theta_1,\theta_2\}& \{\theta_2,\theta_2\}
    \end{pmatrix}.
\end{align}

All these demands can be easily achieved in the minimal way by choosing $G_a=(C_i,p_i)$, so that $V_a^b=0$ and $U_{ab}^c$ can be read from \eqref{eq:Ci_bracket}. Quantization is then obtained as follows. First, we introduce two canonical pairs of Grassman odd ghost fields $(c^i, \bar b_i)$ and $(\bar{c}^i, b_i)$. Using them, a path integral for the theory can be constructed 
\begin{align}\label{eq:path_int_np}
    {\cal Z}=\int [D\gamma_{ij}][D\pi^{ij}][DN][Dp_0][DN^i][Dp_i][Dc^i][Db_i][D\bar{c}^i][D\bar{b}_i] \ \delta(\theta_1)\delta(\theta_2)\sqrt{\det\mathfrak{M}}\ e^{iS},
\end{align}
where the total action is defined in the following manner
\begin{align}
    S=\int dt d^dx\left[\pi_{ij}\partial_t \gamma_{ij}+p_0 \partial_t N + p_i\partial_t N^i + \bar b_i \partial_t c^i +b_i \partial_t \bar{c}^i -{\cal H}_\Psi\right],
\end{align}
with ${\cal H}_\Psi$ the gauge-fixed Hamiltonian, obtained by adding to ${\cal H}_0$ a BRST exact term fixing the first class constraints. The choice that reproduces the gauge fixing sector of pHG in \cite{Barvinsky:2015kil} for the spatial diffeomorphisms is
\begin{align}
    {\cal H}_\Psi={\cal H}_0+{\cal H}_i N^i + \bar b_i b^i - \bar b_i\left(N^j \partial_j c^i+N^i \partial_j c^j\right)+p_i \chi^i + \bar c_i \{\chi^i , {\cal H}_j\}_Dc^j + \bar c_i \frac{\delta \chi^i}{\delta N^j}b^j,
\end{align}
where $\chi^i$ is the gauge fixing choice. Around flat space it can be written in terms of the perturbation of the spatial metric $h_{ij}$. Notice that, particularities of the Dirac brackets aside, we can identify the usual terms in the quantization of gauge theories with primary constraints here, albeit written in BRST formalism \cite{Barnich:2000zw}.

The measure of the path integral \eqref{eq:path_int_np} includes explicitly the secondary constraints, in the form of two delta functions, together with $\sqrt{\det \mathfrak M}$. The constraint $\theta_2=0$ can be applied algebraically directly in the path integral by setting $p_0=0$, but $\theta_1$, involving the rest of the fields and their derivatives, needs to be promoted to the action with an auxiliary field
\begin{align}
    \delta(\theta_1)=\int [D{\cal A}]\ \exp \left(i \int dt d^dx\ {\cal A}\theta_1\right).
    \end{align}

The same has to be done with the contribution of $\mathfrak M$. Luckily, due to the diagonal structure in \eqref{eq:M_matrix}, the factor simplifies to $\sqrt{\det \mathfrak M}=\det \{\theta_1,\theta_2\}$. This is then added to the path integral by yet another one pair of ghost fields
\begin{align}
    \det\{\theta_1,\theta_2\}=\int [D\bar \eta][D\eta] \ \exp\left(i \int dt d^dx\ \bar\eta \frac{\delta \theta_1}{\delta N}\eta\right),
\end{align}
where we have used the fact that $\det\{\theta_1,\theta_2\}$ greatly simplifies due to the trivial form of $\theta_2$.

The authors of \cite{Bellorin:2022qeu} then proceed to expand around flat space, adding to \eqref{eq:flat_space_perturbation} a perturbation of the lapse $N=1+n$. They propose a choice for $\chi^i$ that reproduces the regular gauge structure found in pHG by \cite{Barvinsky:2015kil}. However, non-regular terms remain in the propagators of several fields. In $D=2$, which is enough for our discussion here, these are
\begin{align}
    \langle {\cal A}(\omega,k){\cal A}(-\omega,-k)\rangle=\langle {\cal A}(\omega,k)n(-\omega,-k)\rangle=\langle \eta(\omega,k)\bar{\eta}(-\omega,-k)\rangle \supset \frac{1}{k^4}.
\end{align}

These are the only propagators which involve fields that are not present in the quantization of the projectable theory -- when done in an equivalent Hamiltonian approach, of course. Although their irregular character is unavoidable -- no choice of $\chi^i$ cancels them, as discussed in \cite{Bellorin:2022qeu} -- their presence is actually harmless. The fact that all propagators and vertices which couple $n,{\cal A},\eta$, and $\bar{\eta}$ with the rest of fields can be derived from variations of $\theta_1$ with respect to $n$ allows to show a very non-trivial cancellation among their contribution to different Feynman diagrams. In short -- see \cite{Bellorin:2022qeu} for details -- all contribution of inner $n$ legs to divergent diagrams is always exactly cancelled by that of ${\cal A}$, rendering all divergences explicitly local. The theory is thus power-counting renormalizable, with superficial degree of divergence given by \eqref{eq:degree_divergence_horava}.

Despite this, a complete proof of renormalizability is not available yet, due to the presence of the second-class constraints, which exclude the full theory from those considered in \cite{Barvinsky:2017zlx}. At this point, there are two possible ways to proceed. A first option is to compute explicitly the renormalization group flow of the theory. This is a lengthy task that is however possible using the propagators and vertices derived from \eqref{eq:path_int_np}. Nevertheless, without an explicit proof of gauge invariance of the effective action, a partial result in an specific gauge is not definitive. A second possible route is then to try to generalize the result of \cite{Barvinsky:2017zlx} to the non-projectable theory, but this is a task that remains unexplored so far. Still, given the quantization scheme of \cite{Bellorin:2022qeu}, the possibility that the full non-projectable \Horava gravity is renormalizable seems closer that ever.


\section{Ho\v rava Gravity at low energies}\label{sec:low_energies}

As we have mentioned before, the presence of the term $a_ia^i$ in the Lagrangian of non-projectable \Horava Gravity implies that the propagation speed of the scalar mode will generically differ from that of the graviton mode -- see \eqref{eq:propagation_speed}, so that Lorentz violations are generically present at all energies. At first sight, this might look like an important problem for \Horava Gravity, due to the strong constraints on Lorentz violating interactions obtained from accelerator physics \cite{Liberati:2013xla}. However, these are restricted to the particle content of the Standard Model (SM), and do not apply to the case of gravitational interactions. Indeed, as we see in what follows, the possibility of observing Lorentz violations at low energies in the gravitational sector can be used to constrain \Horava gravity \emph{but not to completely rule it out} \cite{Yagi:2013ava,Yagi:2013qpa,Gupta:2021vdj}. A large region of the parameter space survives observational tests from Big Bang nucleosynthesis \cite{Carroll:2004ai,Audren:2014hza}, from the absence of gravitational Cerenkov radiation \cite{Elliott:2005va}, and in particular from the propagation and emission of gravitational waves from compact objects \cite{EmirGumrukcuoglu:2017cfa,LIGOScientific:2017ync}, which set the most restricting bounds. Beyond this, we will also discuss here a possible mechanism to protect the species in the SM from the percolation of Lorentz violations due to loop corrections \cite{Pospelov:2010mp}, and we will explore some interesting ideas to validate the possibility of pHG as a phenomenological viable theory -- in particular, the appearance of a dark matter contribution to cosmological dynamics without the need of extra fields \cite{Mukohyama:2009mz}.

\subsection{Einstein-Aether Gravity vs. Khronometric Gravity}\label{sec:EA_k}

Before discussing the low energy phenomenology of \Horava Gravity, let us take a detour and set up a more comfortable formalism for the computations that follow, by introducing Einstein-Aether (EA) gravity \cite{Jacobson:2000xp}, with action
\begin{align}\label{eq:action_EA}
    S_{\rm EA}=\frac{1}{16\pi G_{\rm AE}}\int d^4x \sqrt{|g|}\left(-\hat R+M_{\m\n}^{\a\b} \hat\nabla_\a U^\m\hat\nabla_\b U^\n+ \varrho (U_\mu U^\mu-1)\right),
\end{align}
where $M^{\a\b}_{\m\n}$ is given by
\begin{align}
    M^{\a\b}_{\m\n}=c_1 g^{\a\b}g_{\m\n}+c_2 \delta^\a_\m \delta^\b_\n + c_3 \delta^\a_\n\delta^\b_\m + c_4 U^\a U^\b g_{\m\n},
\end{align}
and $c_i$ are dimensionless coupling constants. Note that we have allowed for a Newton constant $G_{\rm AE}$ which might differ from that entering onto Newton's law.

We highlight several points here. First, we have returned to a $d$-dimensional formalism, abandoning the ADM decomposition of the metric, and set $d=4$ explicitly. Second, we have coupled to the Einstein-Hilbert action a time-like vector $U^\m$ -- the \emph{aether} -- with unit norm, enforced by the Lagrange multiplier $\varrho$. Finally, we have written the most general action involving $g_{\m\n}$ and $U^\m$ up to second derivatives.

The unit norm condition for $U^\m$ can be interpreted as an explicit breaking of Lorentz invariance, so that the integral lines of $U^\m$ describe a preferred threading defining a time direction $\partial_t\equiv U^\m \partial_\m$. EA gravity is thus the most general Lorentz violating gravitational action up to second order in derivatives. Around flat space-time it propagates a tt graviton, a transverse vector and a scalar mode \cite{Jacobson:2004ts}, with speeds
\begin{align}
    c_{tt}^2=\frac{1}{1-c_{13}},\quad c_{v}^2=\frac{c_1-\frac{1}{2}c_1^2+\frac{1}{2} c_3^2}{c_{14} (1-c_{13})},\quad c_{s}^2=\frac{c_{123}(2-c_{14})}{c_{14}(2(1+c_2)^2-c_{123}(1+c_2+c_{123})},
\end{align}
where we have introduced the notation $c_{ij\dots k}=c_i+c_j+\dots c_k$.

Being the most general Lorentz violating theory with up to two derivatives, it is clear that it must include the low energy limit of \Horava Gravity -- defined by letting $M_*\rightarrow \infty$ in \eqref{eq:np_action}, thus retaining only second derivatives -- as a particular case. This can be seen more clearly by writing the Lagrangian in terms of irreducible representations of the $SO(3)$ Lorentz subgroup that leaves the aether invariant \cite{Jacobson:2013xta}, so that
\begin{align}
   \hat \nabla_\m U_\n=-\frac{1}{3}\theta \mathfrak{h}_{\m\n} + \sigma_{\m\n}+\omega_{\m\n}+U_\m a_\n,
\end{align}
where $\mathfrak{h}_{\m\n}=g_{\m\n}-U_\m U_\n$ is the projector onto the hyperspace orthogonal to $U^\m$, and $\theta$, $\sigma_{\m\n}$, $\omega_{\m\n}$ and $a_\m$ are respectively the expansion, shear, twist (or vorticity), and acceleration of the aether, defined as
\begin{align}
    \theta=\hat\nabla_\m U^\m,\quad \sigma_{\m\n}=\hat\nabla_{(\m}U_{\n)}+a_{(\m}U_{\n)}-\frac{1}{3}\theta \mathfrak{h}_{\m \n},\quad \omega_{\m\n}=\hat\nabla_{[\n}U_{\m]}+a_{[\m}U_{\n]},\quad a^\m=U^\n\hat\nabla_\n U^\m.
\end{align}
These can be used to rewrite the Lagrangian in a simpler form
\begin{align}\label{eq:action_EA_twist}
    S_{\rm EA}=-\frac{1}{16\pi G_{\rm AE}}\int d^4x \sqrt{|g|}\left(\hat{R}+\frac{c_\theta}{3}\theta^2 + c_\sigma \sigma^2+c_\omega \omega^2 + c_a a^2\right),
\end{align}
where 
\begin{align}
    c_\theta=c_{13}+3c_2,\quad c_\sigma=c_{13},\quad c_\omega=c_1-c_3,\quad c_a=c_{14}. 
\end{align}

Notice that previously we said that the aether defines a preferred \emph{threading} in space-time, but not a foliation, which is the main ingredient of \Horava gravity. The condition sine qua non for its existence is that the vector $U^\m$ is hypersurface orthogonal -- thus being the normal vector to the leafs of the foliation. In that case, by Frobenius theorem, the twist must vanish $\omega_{\m\n}=0$, so that the aether can always be written as the gradient of a scalar
\begin{align}
    U^\m=\frac{\hat{\nabla}^\m T}{\sqrt{\hat{\nabla}^\m T\hat{\nabla}_\m T}}.
\end{align}
Due to this choice, the vector mode within $U^\m$ is eliminated from the spectrum, so that only the transverse traceless and the scalar degrees of freedom remain.

In this form, the aether is left unchanged under a monotonous reparametrization $T\rightarrow T'(T)$, which implies that $T$ thus defines a preferred time direction. Fixing the gauge $T=t$, and opening up the operators in the action \eqref{eq:action_EA_twist} by means of an ADM decomposition in the foliation orthogonal to $U^\m$, one precisely finds \eqref{eq:np_action} in the limit $M_*\rightarrow \infty$, with the identifications \cite{Blas:2010hb}
\begin{align}\label{eq:coupling_relations}
    \frac{G}{G_{\rm AE}}=\eta=\frac{1}{1-c_{13}},\quad \lambda=\frac{1+c_2}{1-c_{13}},\quad \alpha=\frac{c_{14}}{1-c_{13}}.
\end{align}
Note in particular that the gravitational coupling of matter to \Horava Gravity and to EA Gravity can in general differ from the GR one, due to the term $\eta R$ in the action. Although normalizing $\eta=1$ is always possible in the absence of sources, this is not the case anymore when matter is coupled to gravitation. We also observe that the parameters $c_1$ and $c_3$ enter \eqref{eq:coupling_relations} only in the combination $c_1+c_3$, as a consequence of hypersurface orthogonality. The vanishing of the twist implies that we can always shift the value of $c_\omega=c_1-c_3$ by adding to the action a term $\omega^2$ to the action, which vanishes on-shell. Only the combination $c_{13}$ remains physically meaningful. This allows to choose yet another new parametrization that has been used sometimes in the literature \cite{Blas:2010hb}
\begin{align}\label{eq:parametrization_abl}
 c_1=c_3,\quad  c_2=\lambda'=(1-\beta')(\lambda-1)-\beta',\quad c_{13}=\beta'=1-\frac{1}{\eta}, \quad c_{14}=\alpha.
\end{align}
In this form, we can encode deviations from GR due to the dynamics of \Horava gravity solely in three parameters $\lambda'$, $\beta'$ and $\alpha$, which vanish in the GR limit $c_i\rightarrow 0$.

This connection between EA Gravity and \Horava gravity was historically found in reverse. In \cite{Blas:2010hb}, the authors restored full diffeomorphism invariance into Ho\v rava gravity by adding the scalar field $T$ as a Stueckelberg field \cite{Ruegg:2003ps}, that they named \emph{khronon}\footnote{From the ancient Greek word $\chi\rho \acute{o} v o \varsigma$ (khronos, time).}. In this way, not only the covariant expression for the action up to second derivatives -- matching the action of EA gravity \eqref{eq:action_EA} with the identification \eqref{eq:coupling_relations} -- can be found, but a rulebook to construct terms in higher derivatives is provided. In some sense, the work of \cite{Blas:2010hb} can also be interpreted as providing a UV completion for a finite region in the parameter space of EA gravity. In what follows, we will use the EA approach to discuss the low energy limit of non-projectable \Horava gravity, hereinafter named \emph{khronometric theory}, since most of the techniques and intuition developed for GR can easily be extended to its case in a somewhat simple way.

\subsection{Constraints on the parameter space} \label{sec:constraints}
The parameter space of \Horava gravity at low energies has been bounded from different considerations and comparison to observational data. A dedicated summary of results can be found in \cite{Gupta:2021vdj,EmirGumrukcuoglu:2017cfa}, but here we reproduce some of the key points that bound the values of $\alpha,\beta'$ and $\lambda'$. 

A first theoretical constraint comes from considering the propagation speeds of the tensor and scalar modes around flat space-times. In the parametrization \eqref{eq:parametrization_abl} these are
    \begin{align}\label{eq:speeds_abl}
        c_{tt}^2=\frac{1}{1-\beta'},\quad c_s^2=\frac{(\lambda'+\beta')(2-\alpha)}{\alpha(1-\beta')(2+3\lambda'+\beta')}.    
    \end{align}
The requirement of absence of ghosts imposes $c_{tt}^2>0$ and $c_s^2>0$. On the other hand, and in order to prevent ultra-high energy cosmic rays to decay onto gravitational modes in a Cerenkov-like cascade \cite{Elliott:2005va}, both speeds have to satisfy $c_i^2>1-{\cal O}\left(10^{-15}\right)$. Finally, the coincident detection of the gravitational wave event GW170817 and of the gamma ray burst GRB 170817A \cite{EmirGumrukcuoglu:2017cfa,LIGOScientific:2017ync} constrains $-3\times 10^{-15}<c_{tt}-1<7\times 10^{-16}$. In turn, all these considerations together imply the tightest constraint on the theory
    \begin{align}
        |\beta'|\lesssim 10^{-15}.
    \end{align}

The persistence of Lorentz violations down to arbitrary low energies can also have an effect on the dynamics of gravitating objects \cite{Bonetti:2015oda,Foster:2005dk,Yagi:2013ava,Yagi:2013qpa,Gupta:2021vdj}. In particular, the scalar degree of freedom responsible of Lorentz violations can induce extra forces that perturb the motion of bodies under the influence of gravity. The general formalism to describe these deviations is the so-called parameterized post-Newtonian (PPN) formalism \cite{will_2018}, in which the metric -- and other field quantities -- are expanded in powers of $v^2/c^2$ -- where $v$ is the characteristic speed of the gravitating system and $c$ is the speed of light --, and of the dimensionless gravitational potential $GM/(c^ 2 r)$, where $M$ is the mass of the source of the gravitational field and $r$ the distance to it. On practice, one expands the metric $g_{\m\n}$ at leading order as\footnote{Here we follow the notation of \cite{Foster:2005dk}.}
\begin{align}
    \nonumber g_{00}&=1-2U +2\beta^{\rm PPN} U^2 +2\xi^{\rm PPN} \Phi_W -(2\gamma^{\rm PPN} +2 +\alpha_3^{\rm PPN} +\zeta_1^{\rm PPN}-2\xi^{\rm PPN})\Phi_1 \\
   \nonumber &-2(3\gamma^{\rm PPN} - 2 \beta^{\rm PPN} + 1 +\zeta_2^{\rm PPN} + \xi^{\rm PPN})\phi_2 -2 (1+\zeta_3^{\rm PPN})\Phi_3 \\
    &- 2(3\gamma^{\rm PPN} + 3 \zeta_4^{\rm PPN}-2\xi^{\rm PPN})\Phi_4 +(\zeta_1^{\rm PPN}-2\zeta^{\rm PPN}){\cal A},\\
    g_{0i}&=\frac{1}{2}\left(4\gamma^{\rm PPN} +3+\alpha_1^{\rm PPN} - \alpha_2^{\rm PPN}+\zeta_1^{\rm PPN}-2\xi^{\rm PPN}\right)V_i +\frac{1}{2}\left(1+\alpha_2^{\rm PPN}+\zeta_1^{\rm PPN}+2\xi^{\rm PPN}\right)W_i,\\
    g_{ij}&=-(1+2\gamma^{\rm PPN} U )\delta_{ij},
\end{align}
where the potentials $U,\Phi_n , {\cal A},V_i$, and $W_i$ are functions of the spatial coordinates defined by the expression
\begin{align}
    F(x)=G_N\int d^3y\ \frac{\rho(y) f}{|x-y|},
\end{align}
with the substitutions $F:f$
\begin{align}
   \nonumber  &U:1,\quad \Phi_1: v_i v_j,\quad \Phi_2:U,\quad \Phi_3 :\Pi , \quad \Phi_4:p/\rho,\\
    &\phi_W: \int d^3z \rho(z)\ \frac{(x-y)_j}{|x-y|^2}\left(\frac{(y-z)_j}{|x-z|}-\frac{(x-z)_j}{|y-z|}\right),\quad {\cal A}:\frac{(v_i(x-y)_i)^2}{|x-y|^2},\\
    \nonumber &V_i: v^i, \quad W_i: \frac{v_j (x_j-y_j)(x^i-y^i)}{|x-y|^2}.
\end{align}

Here, $\rho, p, \Pi$ and $v_i$ are the energy density, isotropic pressure, internal energy density, and three-velocity of the sources of the gravitational field, described by a fluid with energy-momentum tensor
\begin{align}
    T^{\m\n}=(\rho +\rho \Pi +p)v^\m v^\n-pg^{\m\n},
\end{align}
with $v^\m$ its four-velocity. $G_N$ is the Newton's constant, defined as the coupling entering into the force between two point particles in the Newtonian limit. In the parametrization \eqref{eq:parametrization_abl} it is given by
\begin{align}
    G_N=G_{\rm AE}\left(1-\frac{\alpha}{2}\right)^{-1}.
\end{align}

One then proceeds to plug this expansion onto the equations of motion coming from \eqref{eq:action_EA}, and to solve them order by order in the potentials. The leading post-Newtonian result then fixes the PPN parameters to
\begin{align}\label{eq:PPN_Horava}
   \nonumber  &\gamma^{\rm PPN}=\beta^{\rm PPN}=1,\\
   \nonumber &\xi^{\rm PPN}=\zeta_1^{\rm PPN}=\zeta_2^{\rm PPN}=\zeta_3^{\rm PPN}=\zeta_4^{\rm PPN}=\alpha_3^{\rm PPN}=0,\\
    &\alpha_1^{\rm PPN}=4\frac{\alpha-2\beta'}{\beta'-1}\\
   \nonumber &\alpha_2^{\rm PPN}=\frac{\alpha_1^{\rm PPN}}{8+\alpha_1^{\rm PPN}}\left(1+\frac{\alpha_1^{\rm PPN}(1+\beta'+2\lambda'}{4(\beta'+\lambda')}\right).
\end{align}

The first two parameters $\beta^{\rm PPN}$ and $\gamma^{\rm PPN}$ are known as the Eddington-Robertson-Schiff parameters, and characterize the non-linearity and spatial curvature produced by gravity. At this order, their value matches that of GR, as it happens with the vanishing parameters in the second line. The only non-vanishing quantities are $\alpha_1^{\rm PPN}$ and $\alpha_2^{\rm PPN}$, which characterize preferred frame effects. Their value can be bounded by precise measurement of the orbit of the Moon through Lunar laser ranging \cite{Muller:2005sr} -- thanks to the mirrors left by the Apollo missions -- and of binary pulsar systems -- such as PSR J1738+0333 \cite{Freire:2012mg}. Comparison of the latter with theoretical predictions is done through perturbations of the Newtonian orbit and the Einstein-Infeld-Hoffman method \cite{Will:2018ont}. In the point-particle approximation for the bodies, this can be computed by defining their action in the standard way, as the integral of their mass along their world-lines
\begin{align}
    S_{\rm pp}=-\int m_A (\gamma_A)\ d\tau_A.
\end{align}
Here $A$ labels the body, with $\tau_A$ its proper time. Deviations from GR are encoded in the dependence of the mass $m_A$ on the parameter $\gamma_A$, which in the case of \Horava Gravity is the product $\gamma_A=(v_A\cdot U)$ of the body's four-velocity with the aether. Then the sensitivities
\begin{align}
    \sigma_A=\left.-\frac{d\log(m_A)}{d \log(\gamma_A)}\right|_{\gamma_A=1},
\end{align}
are defined through a Taylor expansion around the GR configuration \cite{Will:2018ont,Gupta:2021vdj}. From here, one then proceeds by noting that the sensitivities of a given body can also be obtained by looking at the asymptotic expression for the metric of the body in motion \cite{Will:2014kxa}. Long story short, the leading contribution proportional to the speed $v$ in the metric contains the sensitivity $\sigma_A$. 

This strategy was put in motion for EA gravity as well as \Horava gravity in \cite{Gupta:2021vdj}, following previous works in \cite{Yagi:2013ava,Yagi:2013qpa}. By constructing numerical models for neutron stars moving with constant linear speed $v$, they extracted the value of the sensitivities, which control in particular the energy loss of a binary system due to emission of GWs \cite{Foster:2006az} 
\begin{align}
\nonumber \frac{\partial_t E_{\rm bin}}{E_{\rm bin}}&=2\left\langle \left(\frac{{\cal G}G \mu m}{r_{12}^3}\right)\left\{\frac{32}{5}({\cal A}_1 + {\cal S}{\cal A}_2 + {\cal S}^2 {\cal A}_3)v_{12}^2 \right\}\right.\\
&\left. +(s_1-s_2)^2\left[ {\cal C}+\frac{18}{5}{\cal A}_3 V^j_{\rm CM}V^j_{\rm CM} +\left(\frac{6{\cal A}_3}{5}+36{\cal B}\right)\left(V^i_{CM}\hat{n}_{12}\right)^2 \right]   \right\rangle,
\end{align}
where we have rescaled the sensitivities of the two bodies $1,2$ to $s_A=\sigma_A/(1+\sigma_A)$, and defined the coefficients
\begin{align}
  \nonumber &{\cal A}_1=\frac{1}{c_{tt}}+\frac{3\alpha ({\cal Z}-1)^2}{2 c_s (2-\alpha)}, \quad {\cal A}_2=\frac{2({\cal Z}-1)}{c_s^3(\alpha-2)},\quad {\cal A}_3=\frac{2}{3\alpha(2-\alpha)c_s^5},\quad {\cal B}=\frac{1}{9\alpha c_s^5(2-\alpha)}\\
  &{\cal C}=\frac{4}{3 c_s^3 \alpha(2-\alpha)},\quad {\cal G}=\frac{G_N}{(1+\sigma_1)(1+\sigma_1)},\quad {\cal S}=s_1\frac{m_2}{m}+s_2\frac{m_1}{m},\quad {\cal Z}=\frac{(\alpha_1^{\rm PPN}-2\alpha_2^{\rm PPN})(1-\beta')}{3(2\beta'-\alpha)},
\end{align}
with $m=m_1+m_2$, $\mu=(m_1m_2)/m$, $V_{\rm CM}$ the velocity of the center of mass of the binary, $\hat{n}$ the vector normal to the orbital plane, and $v_{12}$ the relative velocity. The active mass $m_A$ is defined through the rest mass $\tilde m_A$ of the bodies by $m_A=\tilde m_A (1+\sigma_A)$.

Using the previous expression and the numerical model for the sensitivities, the preferred frame parameters $\alpha_1^{\rm PPN}$ and $\alpha_2^{\rm PPN}$ can be bound to be $|\alpha_1^{\rm PPN}|\lesssim 10^{-5}$ and $|\alpha_2^{\rm PPN}|\lesssim 10^{-7}$ \cite{Gupta:2021vdj}. Further constraints through fine measurement of the orbital motion should be possible, but theoretical computations are complicated. In particular, in \cite{Gupta:2021vdj} it was shown that the perturbation of the orbital parameters due to preferred frame effects depends not only on the sensitivities but also on their derivatives -- second order corrections to $m_A$. Computing them through numerical models of neutron stars is however beyond the state-of-the-art of the topic.

Taking into account the previous constraint in $\beta'$, and that thus $|\lambda'|\gg |\beta'|$, the expressions in \eqref{eq:PPN_Horava} can be simplified to
\begin{align}\label{eq:bounds}
    4|\alpha|\lesssim 10^{-5},\quad \left|\frac{\alpha}{\alpha-2}\right|\left|1-\alpha \frac{1+2\lambda'}{\lambda'}\right|\lesssim 10^{-7}.
\end{align}

These conditions can be satisfied in two manners. For $|\lambda'|\gg 10^{-7}$, we obtain $|\alpha|\lesssim 10^{-7}$. On the other hand, we can also have $\lambda' \sim \alpha/(1-2\alpha)$ and $|\alpha|\lesssim 0.25\times 10^{-5}$. This latter option implies that all parameters are very small and thus the dynamics of the theory is always very close to GR. On the other hand, the former option, while still constraining $\alpha$ and $\beta'$ to small values, leaves $\lambda'$ essentially unconstrained.

Possible independent tests for $\lambda'$ are scarce. In \cite{Franchini:2021bpt}, it was shown that the dynamics of gravitational collapse, as well as the perturbations of spherically symmetric metrics -- which would constitute the basis to compute the quasinormal mode spectrum controlling the ringdown dynamics in \Horava gravity -- are indistinguishable from those of GR, the equations of motion being equivalent in these two particular cases. This statement goes beyond this result, according to early analysis of the constraint structure of the theory when $\alpha=\beta'=0$. In particular, in \cite{Loll:2014xja} it is claimed that the theory with only non-vanishing $\lambda'$ is indistinguishable from GR in asymptotically flat space-times. The only possibility to bound $\lambda'$ seems to be to look at a situation without flat asymptotics. The obvious choice is then to turn to Cosmology \cite{Audren:2014hza}. Imposing a Friedmann-Robertson-Walker form for the metric leads to a standard Friedmann equation\footnote{Although with subtleties, see section \ref{sec:sec:DM_integration} later.} for the Hubble parameter $H$
\begin{align}
    H^2=\frac{8\pi}{3}G_{C}\rho,
\end{align}
albeit with a different definition for the Newton's constant
\begin{align}
    \frac{G_N}{G_C}=\frac{2+\beta'+3\lambda'}{2-\alpha}\sim 1+\frac{3\lambda'}{2},
\end{align}
where in the last step we have used the previous bounds for $\alpha$ and $\beta'$. In order to correctly predict the abundance of primordial elements during Big Bang nucleosynthesis, the previous ratio has to satisfy $|G_C/G_N-1|\lesssim 1/8$, which imposes -- together with the requirement of absence of ghosts -- a relatively weak bound $0\lesssim \lambda' \lesssim 0.1$. In the particular case that $\alpha=\beta'=0$ exactly, this bound can be improved to $0\lesssim \lambda' \lesssim 0.01$ -- see \cite{Afshordi:2009tt}. Finally, there seems to be indications that a theory with $\alpha\neq 0$ and $\beta'\neq 0$ is inconsistent, as moving black holes seems to develop singularities at the position of their horizons \cite{Ramos:2018oku}. However, more recent results have questioned this conclusion\footnote{See talk by A. Kovachik, "Slowly Moving Black Holes in Hořava Gravity: Revisited", at the IFPU Focus program: Lorentz violations in gravity; held in Trieste in July 2023.}.


\subsection{Black holes in \Horava gravity}\label{sec:black_holes}

The fact that a certain amount of Lorentz violation always survives down to low energies has strong implications for the structure of black holes in \Horava gravity. A priory, one could even question the existence of black holes as causally disconnected compact regions of space-time. The reason is simple -- for non-vanishing $\alpha, \beta'$ and $\lambda'$, the speed of the scalar mode \eqref{eq:speeds_abl} is always super-luminal. Hence, perturbations of this field can always enter and exit the usual event horizon, as defined in GR. The same is true for matter fields. Even if their IR speeds are sub-luminal, the presence of higher derivative terms leads to modified dispersion relations of the form $\omega^2=c_A^2 k^2 + b_A^2 k^4+\dots$ for some couplings $c_A$ and $b_A$, and can induce momentum dependent speeds larger than one. However, this naive reasoning fails when applied to the specific case of \Horava gravity, as we will see in what follows. 

For the sake of a more concrete discussion, let us work in the framework of EA gravity and introduce a four-dimensional spherically symmetric and static space-time in Schwarzschild coordinates
\begin{align}\label{eq:bh_metric}
    ds^2=F(r)dt^2 -\frac{B(r)^2}{F(r)}dr^2-r^2 dS_2^2,
\end{align}
where $dS_2^2= d\theta^2 + \sin^2 \theta d\varphi^2$ is the metric of the two-dimensional sphere, and $F(r)$ and $B(r)$ are undetermined functions of the radial coordinate. This metric enjoys a time-like Killing vector $\chi^\m=(1,0,0,0)$ defining staticity, and with norm $|\chi|^2=F(r)^2$, which will be important in our discussion later. The corresponding aether satisfying the isometries of \eqref{eq:bh_metric} takes the form
\begin{align}
    U_\m dx^\m=\frac{1+F(r)A(r)^2}{2A(r)}dt+\frac{B(r)}{2A(r)}\left(\frac{1}{F(r)}-A^2(r)\right)dr, 
\end{align}
and has a single degree of freedom $A(r)$, due to staticity, spherical symmetry, and the unit norm condition $U_\m U^\m=1$. The specific form of the components of $U^\m$ has been chosen for computational convenience. We will also assume asymptotic flatness -- implying $F(r)=B(r)=A(r)=1$ for $r\rightarrow \infty$. As noted by \cite{Eling:2006ec}, this aether is automatically hypersurface orthogonal -- satisfying $\omega_{\m\n}=0$. Hence all spherically symmetric solutions of EA gravity are automatically solutions of khronometric gravity at low energies. For arbitrary space-time isometries instead, only the reverse is true. 

The specific shape of the functions $F(r),B(r)$ and $A(r)$ must be obtained by solving the equations of motion coming from variation of the action \eqref{eq:action_EA}, although solutions exist everywhere in the parameter space \cite{Barausse:2011pu}. However, they can only be obtained numerically. Only in two regions, identified in \cite{Berglund:2012bu}, the solution can be specified analytically (in both cases $r_0$ is an integration constant)
\begin{itemize}
    \item \bf{Case 1:} $c_{14}=0$, $(\alpha=0)$
\begin{align}
   \nonumber  &F(r)=1-\frac{r_0}{r}-\frac{c_{13}r_{\rm AE}^4}{r^4},\quad  B(r)=1,\\
    &A(r)=\frac{1}{F(r)}\left(-\frac{r_{\rm AE}^2}{r^2}+\sqrt{F(r)+\frac{r_{\rm AE}^4}{r^4}}\right),\quad r_{\rm AE}=\frac{r_0}{4}\left(\frac{27}{1-c_{13}}\right)^{\frac{1}{4}}.
\end{align}
\item \bf{Case 2:} $c_{13}+c_2=0,\quad (\beta'+\lambda'=0)$
\begin{align}
    \nonumber &F(r)=1-\frac{r_0}{r}-\frac{{\cal R} (r_0+{\cal R})}{r^2},\quad B(r)=1\\
    & A(r)=\frac{1}{1+\frac{{\cal R}}{r}},\quad {\cal R}=\frac{r_0}{2}\left(\sqrt{\frac{2-c_{14}}{2(1-c_{13})}}-1\right).
\end{align}
\end{itemize}
Note that in light of the bounds discussed in section \ref{sec:constraints}, case 2 is compatible with observations only if $\lambda'=0$, which then implies $\alpha'=0$ from \eqref{eq:bounds}, reducing the theory to GR. Case 1 however remains compatible with non-trivial values of the couplings in the Lagrangian.

Since the aether is hypersurface orthogonal, it is natural to adapt our chart of coordinates to that of Eulerian observers flowing with the foliation orthogonal to $U^\m$. This is achieved by letting
\begin{align}\label{eq:change_tau}
    d\tau=dt+\frac{U_r}{U_t}dr,
\end{align}
which aligns the time direction with the preferred time $\tau$. Although non-mandatory, it is also convenient to perform an extra spatial diffeomorphism to arrive to the diagonal gauge for the metric
\begin{align}\label{eq:change_rho}
    d\rho=dt+\frac{S_r}{S_t}dr,
\end{align}
where $S^\m$ is the unit vector orthogonal to the aether
\begin{align}\label{eq:S_vector}
    S_\m dx^\m =-\frac{1-F(r)A(r)^2}{2A(r)}dt-\frac{B(r)(1+F(r)A(r)^2)}{2A(r)F(r)}dr.
\end{align}

After this, the metric takes an explicit ADM form, with 
\begin{align}
    N=\frac{1+F(r)A(r)^2}{2A(r)},\quad N^i=0,\quad \gamma_{ij}dx^i dx^j=\left(\frac{1-F(r)A(r)^2}{2A(r)}\right)d\rho^2 +dS_2^2.
\end{align}

Note that this chart of coordinates has a problem when $N=0$, which corresponds to $1+F(r)A(r)^2=0$ and to the vanishing of $U_t$ at a radial point $r_U$. From \eqref{eq:change_tau} we see that this point lays at finite $r$ but it is mapped to $\tau\rightarrow +\infty$, signaling that the foliation cannot be globally extended in a smooth way beyond this point. Also, the product $(\chi\cdot U)$ vanishes at this point, which implies an acausal surface such that interior and exterior cannot be casually connected \cite{Bhattacharyya:2015gwa}. This is hence a trapping surface for all trajectories regardless of their propagation speed, reason why it is known as \emph{universal horizon} (UH). Note that its position always lays behind the Killing horizon $F(r)=0$ -- otherwise $(\chi\cdot U)=0$ would not be possible, since $U^\m$ is time-like everywhere -- and, more importantly, that this coordinate problem cannot be removed by a change of chart -- unlike what happens with the usual horizons in GR -- since once in the preferred frame, the symmetry group is restricted to FDiff \eqref{eq:fdiff}.

Note also that since $N=0$, and in order to keep all metric functions and the action smooth at this point, $N$ must reverse sign when crossing the UH, a fact that is again unavoidable by a FDiff transformation \cite{DelPorro:2022kkh}. Hence, the interior of the UH is causally reversed with respect to its exterior, enjoying a different time orientation. This fact is important in ensuring a correct behavior of a quantum field theory living in the environment of these solutions, and in providing a reservoir of negative energy for particle production \cite{Schneider:2023cuo}.

Once the existence of the UH is established, the standard lore of general relativistic black holes can then be translated to these solutions almost straightforwardly. We simply need to replace the usual implications of the presence of the event horizon to the UH. It is however important to remark here that metrics of the form \eqref{eq:bh_metric} are solutions to the low energy part of the gravitational action of Ho\v rava gravity only, where $M_*$ has been formally taken to arbitrarily large energies. It is unknown if the UH is actually a stable surface or simply a transient feature that disappears once the full action is considered.

All the results that we have discussed here for spherically symmetric black holes are equally valid for EA gravity and for \Horava gravity, since spherical symmetry and staticity automatically imply $\omega_{\m\n}=0$, as we have mentioned. However, this property will not translate to more general space-times and, in particular, it does not hold in the case of axially-symmetric solutions -- i.e rotating black holes. This inequivalence has actually led to controversial results in the literature. The search for slowly rotating black holes -- those which are perturbative in deviations from sphericity -- was done in \cite{Barausse:2015frm,Barausse:2012qh}, where the leading order solutions were found, displaying large differences with their counterparts in EA Gravity. Slowly rotating solutions in the latter theory have $\omega_{\m\n}\neq 0$, which implies that no preferred foliation exists -- instead, there is only a preferred frame defined locally at every space-time point. Due to this, it is not possible to fulfil the UH condition $(\chi\cdot U)=0$. In contrast, slowly rotating solutions in \Horava Gravity seem to reproduce the properties of static ones. At leading order in rotation (${\cal O}(\epsilon)$), and assuming asymptotic flatness, the metric and aether vector read
\begin{align}
    &ds^2=F(r)dt^2 -\frac{B(r)}{F(r)}-r^2 dS_2^2 + \epsilon r^2 \sin^2 \theta \ \Omega(r,\theta) dt d\varphi + {\cal O}(\epsilon^2),\\
    &U_\m dx^\m=\frac{1+F(r)A(r)^2}{2A(r)}dt+\frac{B(r)}{2A(r)}\left(\frac{1}{F(r)}-A^2(r)\right)dr + {\cal O}(\epsilon^2),
\end{align}
where $F(r),B(r)$ and $A(r)$ are given by the static solution. Note in particular that the aether is not corrected at leading order, and hence one expects to find a UH in the same position as for the static case. The only new function here is the rotating factor $\Omega(r,\theta)$, which satisfies
\begin{align}
    \Omega(r,\theta)=\Omega(r)=-12J\int_{r_H}^r \frac{B(\rho)}{\rho^4}d\rho +\Omega_0.
\end{align}
Both $J$ and $\Omega_0$ are integration constants. The former is identified with the spin of the black hole by an asymptotic expansion at large distances, while the second can be absorbed by a coordinate change $\varphi'=\varphi - \Omega_0 t/2$.

Only recently, \cite{Adam:2021vsk} performed an exhaustive search of rotating black holes within EA gravity. They used the approach of \emph{painting} the aether on top of a rotating GR solution, which drastically reduces the complexity of the computation. In their results, the authors of \cite{Adam:2021vsk} find a similar picture to the one that we have described. The aether is never twist-free in their solutions. Although the existence of a rotating solution in the \Horava limit is conjectured in most of the works in the topic, its explicit computation is still an open problem.

\subsection{Percolation of Lorentz violations to the Standard Model}\label{sec:percolation}

There is a subtle issue that we have ignored so far in our discussion. Even in the case in which Lorentz violating effects in the gravitational sector remain small, bounded by the results in section \ref{sec:constraints}, we must worry about their percolation to the SM fields, since deviations from Lorentz invariance in the matter sector are strongly constrained -- see e.g. \cite{Liberati:2013xla} and references therein. This is worrisome, because corrections to the propagation speed of the different species in the SM, as well as many other physical effects, are controlled by operators of dimension four, hence carrying dimensionless coupling constants. Loop corrections to these, regardless of their origin, can only be logarithmic and thus we could expect them to be ${\cal O}(1)$, clearly larger than what experiments have measured. This argument is the reason why Lorentz violating theories are typically disregarded, and could in principle be used to disfavor the possibility of Ho\v rava gravity providing a realistic model of Nature. However, in the specific case of the latter, there are subtleties that can provide a way out from this conclusion \cite{Pospelov:2010mp}.

In order to be concrete, let us assume a purely Lorentz invariant SM sector with up to dimension four operators, coupled to Ho\v rava gravity, which at low energies reduces to \eqref{eq:action_EA} in a particular point of the parameter space, as discussed in section \ref{sec:EA_k}. Importantly, note that this action is second order in derivatives, with the only dimensionful coupling being the Planck mass $M_P\sim G_{\rm AE}^{-1/2}$. Thus, a simple power-counting argument shows that corrections to dimension four operators induced by gravitational contributions -- those with loops containing $g_{\m\n}$ and $U^\m$ -- will take the form
\begin{align}
    \delta {\cal O}_4\propto \frac{\Lambda^2}{M_P^2}\log \left(\frac{\mu}{\Lambda}\right),
\end{align}
where $\Lambda$ is the cut-off of the theory and $\mu$ is an IR subtraction point -- which can be also thought as the renormalization scale.

If the theory were valid up to $M_P$, as we would expect if gravity is given by GR, then we encounter the problem mentioned before. The cut-off becomes $\Lambda\sim M_P$ and the correction is of order ${\cal O}(1)$, too large to agree with experiments. However, in Ho\v rava gravity, new interactions, controlled by $M_*$, will develop before $M_P$ is reached. In that case, we have $\Lambda\sim M_*$, and the correction is instead of order ${\cal O}\left(M_*^2/M_P^2\right)$. Taking into account the values in \eqref{eq:value_Mstar}, this is
\begin{align}
   10^{-16}<\frac{M_*^2}{M_P^2}<10^{-8},
\end{align}
which is small enough to bypass all experimental constraints. Thus, if the right hierarchy between scales is satisfied, Ho\v rava gravity can provide a way to achieve naturally small deviations from Lorentz invariance in the matter sector.

As a final note, let us also point that even if this is it not the case, there are other ways to protect Lorentz invariance in the SM. One of them is supersymmetry (SUSY), which forbids all Lorentz violating operators of dimension four \cite{GrootNibbelink:2004za}, immediately solving the conundrum.

\subsection{Dark matter as an integration constant}\label{sec:sec:DM_integration}

Let us do a final detour in this section and go back to the projectable version of Ho\v rava gravity, introduced in subsection \ref{sec:pHG}. Let us in particular focus on the problem of building up a cosmological solution, following the ideas by \cite{Mukohyama:2009mz,Mukohyama:2010xz}. For simplicity, we introduce a flat FRW ansatz for the metric
\begin{align}
    ds^2=-dt^2 + a(t)^2 d\Vec{x}^2,
\end{align}
which not only has $N=1$, but also implements the gauge $N^i=0$ for spatial diffeomorphisms.

In contrast to the GR case, here the set equations of motion reduces to a single independent equation
\begin{align}\label{eq:FRWHG}
    -\frac{3\lambda-1}{2}\left(2 \dot{H}+3H^2\right)=8\pi G P,
\end{align}
where $P$ is the pressure of matter coupled as a perfect fluid, and $H=\dot{a}/a$ is the Hubble expansion rate, as usual. Notice that higher derivative terms do not contribute to the equation due to spatial flatness. While in GR the equivalent equation to \eqref{eq:FRWHG} can be derived from the Friedmann equation together with the conservation equation for matter, here it is not the case anymore. First, there is not Friedman equation, since this is derived from the energy constraint, which is absent in pHG due to the projectability condition on the lapse. Second, while we could expect matter to follow the relativistic conservation equation at low energies if Lorentz invariance is restored, in general this should be corrected via Lorentz violating operators, that we generically denote as $\hat{Q}$, so that we have
\begin{align}\label{eq:conservation_rho}
    \dot{\rho}+3 H (\rho+P)=-\hat{Q},
\end{align}
where $\rho$ is the matter density. Whenever $\hat{Q}=0$ we thus recover the relativistic version of the theory.

In GR, the Friedman equation is differentiated in time in order to arrive to the equivalent of \eqref{eq:FRWHG}. Here, we can instead integrate \eqref{eq:FRWHG}, thus arriving to an effective Friedman equation reading
\begin{align}\label{eq:Friedman}
    \frac{3(3\lambda-1)}{2}H^2=8\pi G\left(\rho+\frac{C(t)}{a^3}\right),
\end{align}
where we have also used \eqref{eq:conservation_rho}, and where
\begin{align}\label{eq:DM_C}
    C(t)=C_0+\int_{t_0}^t \hat Q(t')a(t')^3 dt',
\end{align}
with $C_0=C(t_0)$ an integration constant and $t_0$ an arbitrary starting time. 

At low energies $\hat{Q}\rightarrow 0$ and hence $C(t)\rightarrow {\rm constant}$. As a consequence, the last contribution in the rhs of \eqref{eq:Friedman} behaves as presureless dust, and thus dark matter (DM), in this limit. Moreover, \eqref{eq:DM_C} provides an interpretation for its origin as the accumulation of Lorentz violating effects in matter through cosmological history \cite{Mukohyama:2009mz}.

Although we have performed this analysis in a flat FRW space-time, the result applies in the general case, as discussed in \cite{Mukohyama:2010xz}. Due to time reparametrization invariance, the total Hamiltonian of the theory must vanish, as we discussed in section \ref{sec:hamiltonian}. In pHG this is equivalent to enforcing the would-be equation of motion of the lapse, which is a function of time only, before setting it to unity
\begin{align}
    \int d^3 x\left(K_{ij}K^{ij}+\lambda K^2 +{\cal V}\right)=0.
\end{align}

Integrating this equation in time always provides an equivalent to the Friedman equation at low energies, with an extra function constant in space appearing due to integration, and taking the role of the DM component in equation \eqref{eq:Friedman}. 

More interesting features can be found in cosmological solutions to Ho\v rava gravity, although there is a lack of modern literature in the topic. In particular, the Lifshitz scaling \eqref{eq:lifshitz_scaling} can provide an exactly scale-invariant spectrum of cosmological perturbations, while cosmological bounces have been found to happen due to the presence of higher derivative in the action. For a more detailed discussion on these topics, see \cite{Mukohyama:2010xz}.


\section{Beyond low energies: the role of higher derivatives}\label{sec:beyond}

Despite the ample phenomenology implied by the persistence of Lorentz violations down to the IR, the main characteristic of \Horava gravity is the presence of higher spatial derivatives, encoded in FDiff invariant operators. These allow not only for superluminal motion, but also for rainbow effects -- in which the propagation speed of a mode depends on its momentum --, and for corrections to the gravitational interaction in environments with large curvatures. A proper treatment and study of these phenomena is not easy though. At energies large enough, the higher derivative terms dominate the dynamics of the theory and cannot be treated as perturbations, requiring instead a full non-linear analysis of the equations of motion, which can contain up to six derivatives in $D=3$. In spite of these obstructions, efforts to study the phenomenology induced by higher derivatives have not been spare. In this section we summarize some recent results, including attempts to regularize singularities in UV complete models \cite{Lara:2021jul}, and to study the dynamics of a toy model of Lorentz violations on top of spherically symmetric space-times \cite{Oshita:2021onq,implicit}. We will also discuss the emission of Hawking radiation by UHs \cite{Michel:2015rsa,DelPorro:2022kkh,DelPorro:2022vqi,Schneider:2023cuo,Herrero-Valea:2020fqa,Ding:2016srk,Cropp:2016gkn}. 

\subsection{Black holes in UV complete \Horava Gravity}\label{sec:UV_BHs}

As we have discussed at the beginning of this review, GR is an EFT -- the low energy limit of an, in principle unknown, theory of Quantum Gravity. Physical systems with low curvature are well-described by perturbative GR, but the validity of the theory breaks down whenever $R\sim G^{-1}$. Most astrophysical situations never exit this range of validity, and hence GR is usually enough to describe observations to a large grade of accuracy. However, there are theoretical situations where a UV completion is necessary. The most prominent of these cases is the existence of singularities occurring in cosmological spaces and in the interior of black holes. The latter are particularly worrisome, because they are the endpoint of any world-line crossing the event horizon, thus unavoidably belonging to the future of any observer falling into the gravitational well. However, singularities exit the range of validity of GR for obvious reasons, and therefore there is a non-negligible chance that its UV completion will actually smooth them so that space-time is regular everywhere. 

This hand waving argument can actually be tested explicitly in the framework of pHG, since it provides us with a theory satisfying all the necessary requirements -- it is renormalizable, and in $D=2$ it exhibits a UV fixed point with asymptotic freedom. Hence, it seems like a suitable laboratory to investigate the fate of singularities within black holes in Quantum Gravity.

In \cite{Lara:2021jul}, the authors studied solutions to the action \eqref{eq:action_pHG} with potential \eqref{eq:V_phg_2}, and exhibiting staticity and circular symmetry, with metric
\begin{align}
    ds^2=(1-F(r)^2)dt^2-2F(r)dt dr -dr^2 - r^2 G(r)^2 d\theta^2.
\end{align}
This corresponds to imposing the projectability condition $N=1$ and defining a shift and spatial metric invariant under rotations\footnote{Note that in $D=2$, any metric is conformally flat and thus it depends on a single function $G(r)$.}.

Although the equations of motion contain up to fourth derivatives of $G(r)$, they are invariant under rescalings $G(r)\rightarrow k G(r)$ with $k$ constant. This suggest a new variable $\Gamma(r)=1/r + G'(r)/G(r)$ which, together with the use of the Bianchi identities implied by spatial diffeomorphisms, allows to simplify the system to a pair of second order differential equations
\begin{align}
 \label{eq:E1}   &E_1=(\lambda-1)\left(F''+F'\Gamma + F \Gamma'\right)+F\Gamma'+F \Gamma^2=0,\\
 \label{eq:E2}   &E_2=\mu \left(8\Gamma \Gamma'' - 4 (\Gamma')^2 + 8 \Gamma^2 \Gamma' -4 \Gamma^4\right)+(\lambda-1)\left(-(F')^2-2F F' \Gamma - F^2 \Gamma^2\right)-2F F' \Gamma + 2 \hat \Lambda,
\end{align}
where we have omitted the argument of the functions for clarity, and a prime denotes differentiation with respect to $r$.

The low energy limit of this action -- akin to taking the GR limit in $D=3$ -- corresponds to $\mu\rightarrow 0$. In this case, the character of $E_2$ changes and becomes a first order equation. This indicates that the transition between the regimes $\mu\neq 0$ and $\mu=0$ cannot be perturbatively continuous. Still, for large distances, where we expect the curvature to be small, such a limit should lead to the right solution. Assuming thus $\mu=0$ and $\hat \Lambda>0$, the system of equations \eqref{eq:E1} and \eqref{eq:E2} can be solved analytically to yield the following solution in parametric form
\begin{align}
   \nonumber &r\sqrt{\Lambda}=\frac{B}{2}\int^X_-\infty \frac{e^{-X'/2}}{\left(\sinh \frac{-X'}{\sqrt{2\lambda-1}}\right)^{3/2}}dX',\\
    \nonumber &rG=G_\infty B \sqrt{\frac{2\lambda-1}{\Lambda}} \frac{e^{X/2}}{\left(\sinh \frac{-X}{\sqrt{2\lambda-1}}\right)^{1/2}},\\
    \label{eq:sol_d2_lc}&F= \pm B \frac{e^{-X/2}}{\left(\sinh \frac{-X}{\sqrt{2\lambda-1}}\right)^{1/2}},
\end{align}
where $B>0$ and $G_\infty$ are integration constants, and $\lambda <5$, which includes the relevant case of the fixed point $\tilde\lambda=15/14$. At large radii $r\rightarrow \infty$, this solution exhibits de Sitter asymptotics with a possible conical defect
\begin{align}
    &r G\sim G_{\infty} r,\\
    &F\sim \pm r\sqrt{\frac{\hat \Lambda}{2\lambda-1}},
\end{align}
where $G_\infty$ denotes the angle deficit whenever $G_\infty \neq 1$. This enjoys a cosmological Killing horizon at $r_{\rm dS}=\sqrt{(2\lambda -1)/\hat \Lambda}$. The case of vanishing cosmological constant can be achieved by a proper vanishing of $B$ and $G_\infty$ altogether with $\hat \Lambda$, in which case local flat space asymptotics -- albeit with a remaining angle deficit, which obstructs global flatness -- is recovered. 

At short distances, corresponding to $r\ll B/\sqrt{\hat \Lambda}$, the solution instead behaves as
\begin{align}
    F=\pm F_0 \left(r\sqrt{\hat \Lambda}\right)^{-\tilde\sigma},\quad G=G_0 \left(r\sqrt{{\hat \Lambda}}\right)^{2\tilde \sigma}
\end{align}
with
\begin{align}
    \nonumber &\tilde \sigma=\frac{\lambda-2+\sqrt{2\lambda-1}}{5-\lambda}>0,\quad F_0=(\sqrt{2}B)^{1+\tilde \sigma}\left(\frac{3-\sqrt{2\lambda-1}}{2\sqrt{2\lambda-1}}\right)^{-\tilde \sigma},\\
    &G_0=G_\infty \sqrt{2\lambda-1}(\sqrt{2}B)^{-2\tilde \sigma}\left(\frac{3-\sqrt{2\lambda-1}}{2\sqrt{2\lambda-1}}\right)^{1+2\tilde\sigma }.
\end{align}

Remarkably, there is a second Killing horizon located at $r_{H}=F_0^{1/\tilde\sigma}/\sqrt{\hat \Lambda}$, which encloses a curvature singularity when $r\rightarrow 0$. Hence, this solution describes a black hole with de Sitter asymptotics from the point of view of an observer sitting at large distances. However, when diving deep into the gravity well and, in particular getting close to the singularity, the growth of the curvature will imply an exit from the regime of validity of the $\mu=0$ reduction. Indeed, higher derivative terms become important at distances $r<\sqrt{\mu} \ll 1/\sqrt{\hat \Lambda}$. 

The authors of \cite{Lara:2021jul} then proceed to perform an exhaustive analysis using numerical methods and boundary layer theory to test the effect of higher derivatives. We reproduce here the latter, and thus perform an expansion in $\epsilon = (\lambda-1)\ll 1$ -- this condition seems to hold close to the fixed point and towards the IR, taking into account the results from the RG flow. Rescaling the coordinate as $\tilde r= r/\sqrt{\epsilon}$ and performing an expansion in $\epsilon$ we find that $\Gamma(r)=r^{-1}\epsilon^{-1/2}\left(1+\epsilon \tilde g(\tilde r)\right)+{\cal O}(\epsilon^2)$ and $F(\tilde r)$ satisfy the following equations
\begin{align}
    &F''+\frac{F'}{\tilde r}-\frac{F}{\tilde r^2}\left(1- \tilde r \tilde g'-\tilde g\right)=0,\\
    &\frac{8\tilde g''}{\tilde r}-\frac{16 \tilde g}{\tilde r^3}-2 F F'=0,
\end{align}
with derivatives now taken with respect to $\tilde r$.

Numerical integration of the former equations by matching the outer asymptotics to \eqref{eq:sol_d2_lc} always show a curvature singularity at the center of the geometry. Hence, \cite{Lara:2021jul} concludes that higher derivatives, even in a UV complete description of gravity, are not enough to resolve curvature singularities, at least within this family of solutions. Of course, this does not mean that the theory is inconsistent. The solutions shown here could simply not be physically relevant. Or even if they are, they could be describing the exterior configuration of a matter collapse, in which case the interior dynamics would be time-dependent and sourced by the matter fields. Extrapolations of these results to $D=3$ could also fail, due to particularities associated to the space-time dimension.


\subsection{A toy model for perturbations}\label{sec:toy}

As we have discussed throughout this text, understanding the dynamical effects introduced by the preferred frame and the presence of higher derivatives in the action of \Horava Gravity is crucial for a proper assessment of the proposal, and its success in UV completing gravitation. These not only are key for the satisfaction of renormalization requirements, but they should also lead to new phenomenology with respect to GR. Nonetheless, the complicated character of the theory in $D=3$ has obstructed this task so far. Not only there are ${\cal O}(50)$ operators in the action, but also the equations of motion show a dispersive nature even in the simplest case, which entails complications for their integration and evolution in time as an initial value problem.

Some success has been achieved instead by looking at a toy model mimicking all the main features of \Horava Gravity, but undressing it of some of its technical complications. That is the Lifshitz scalar field, heavily used in studies of quantum phase transitions in various strongly correlated electron systems \cite{Ardonne:2003wa}, whose action in covariant form is
\begin{align}\label{eq:Lifshitz_action}
    S_{L}=\frac{1}{2}\int d^4x \sqrt{|g|} \left(\partial_\m \phi \partial^\m \phi + \sum_{i=2}^z \frac{\alpha_{2i}}{\Lambda^{2z-2}}\phi (-\Delta)^i \phi\right),
\end{align}
where $2z$ is the maximum number of spatial derivatives, and $\Lambda$ takes here the purpose of signaling the scale at which Lorentz violations become effective. The operator $\Delta = -\mathfrak{h}^{\m\n}\nabla_\m\nabla_\n$ is the Laplace operator onto the hypersurfaces orthogonal to the aether $U^{\m}$. The total metric $g_{\m\n}$ can thus be decomposed along $\tau$ and its orthogonal co-dimension one hypersurfaces by letting $g_{\m\n}=\mathfrak{h}_{\m\n}+U_\m U_\n$. The action \eqref{eq:Lifshitz_action} can be thought as a toy model for the dynamics of the higher derivatives found in Ho\v rava gravity, but also as the action for a matter scalar field coupled to the latter.

Around flat space, the equations of motion from \eqref{eq:Lifshitz_action} indeed lead to a modified dispersion relation
\begin{align}\label{eq:disp_rel}
    \omega^2=k^2+\sum_{i=2}^z \frac{\alpha_{2i}}{\Lambda^{2z-2}} k^{2i},
\end{align}
with the same dispersive character as those displayed by the degrees of freedom of Ho\v rava gravity. 

Let us note an important feature at this point. Due to the covariant character of \eqref{eq:Lifshitz_action}, the number and character of the solutions to its equations of motion must be the same for all charts of coordinates. However, only when the preferred frame is chosen as a coordinate chart -- also referred sometimes as choosing the unitary gauge for the aether, corresponding to $U_\m dx^\m=N d\tau$ -- the equations of motion are explicitly second order in time derivatives, displaying the ghost-free character of the theory. Any other choice of time coordinate which is not connected to $\tau$ via a FDiff transformation \eqref{eq:fdiff} will lead to higher time derivatives that complicate the analysis, although the theory will eventually remain ghost-free. Still, and even though time evolution will indeed be much more complicated in any other chart of coordinates which is not the preferred one, there are cases in which such a choice can prove convenient for the analysis at hand.

One of these cases corresponds to space-times with time-like Killing vectors. In particular, we will focus from now on in spherically symmetric space-times of the form \eqref{eq:bh_metric}, on top of which we will study the dynamics of the Lifshitz field \eqref{eq:Lifshitz_action}, following the conventions in \cite{Michel:2015rsa,Schneider:2023cuo,DelPorro:2022vqi,Cropp:2013sea}. Hereinforward we will thus suppress the existence of the angular coordinates, working with a spherical configuration for $\phi$. 

The existence of the time-like Killing vector $\chi^\m$ implies conservation of the Killing energy $\Omega$, which is constant along time evolution in the Schwarzschild time, once fixed at a boundary. This is not the case for any other ``energy", and in particular it is not for the energy $\omega(t,r)$ associated to the preferred frame coordinate $\tau$. The same is true for the preferred ``momentum" $k(t,r)$ associated to the always space-like coordinate $\rho$ orthogonal to $\tau$, both defined in \eqref{eq:change_tau} and \eqref{eq:change_rho}. Even though they are not constant, these two quantities still satisfy a dispersion relation, which in the eikonal approximation reduces to \eqref{eq:disp_rel}, as it can be confirmed from \eqref{eq:Lifshitz_action} by inserting a WKB ansatz into the equations of motion.

The space of solutions of \eqref{eq:Lifshitz_action} is in general quite complicated, due to the dispersive character of the equations. However, we can obtain a fair amount of information about the dynamics of the field by retaining the eikonal approximation and solving \eqref{eq:disp_rel} by introducing the relation between the Killing and preferred energies
\begin{align}
    \omega=-\frac{k (S\cdot \chi)+\Omega }{(U\cdot \chi)},
\end{align}
where $S^\m$ is defined in \eqref{eq:S_vector}. This allows to regard \eqref{eq:disp_rel} as an algebraic equation for $k$. 

From now on we will take the EFT point of view for action \eqref{eq:Lifshitz_action} and thus retain only the first possible Lorentz violating contribution to the action, corresponding to $z=2$. In this case \eqref{eq:Lifshitz_action} has in general four solutions for $k$. Imposing boundary conditions so that only those with real $k$ are accepted, this leads to only two possible solutions at large radii, where the geometry is asymptotically flat and $U^\m \propto \chi^\mu$, so that $t\sim \tau$. These are given by
\begin{align}
    \omega_{\infty}=\pm \sqrt{k^2+\frac{\alpha_4 k^{4}}{\Lambda^2}},
\end{align}
and reduce to the general relativistic solutions in the decoupling limit $\Lambda\rightarrow \infty$.

On the other hand, close to the UH -- located at $N=0$ --, there are four real solutions at positive $\Omega$. Although for $z=2$ they can be computed explicitly, their form is not very transparent. A better understanding can be attained by using boundary layer theory \cite{schlichting00:BLT}. Hence, we change variables to $k=\zeta p$ and look for values of $\zeta$ which balance different terms in \eqref{eq:disp_rel}, finding only two possible choices that lead to non-vanishing solutions at leading order in large $\Lambda$
\begin{itemize}
    \item Soft modes $\zeta=\Lambda$
    \begin{align}\label{eq:soft_modes}
        p=-\frac{ \Omega}{(S\cdot \chi)},\quad \omega=\pm\sqrt{k^2+\frac{k^4}{\Lambda^2}}.
    \end{align}
    \item Hard modes $\zeta=\frac{(S\cdot \chi) \Lambda}{(U\cdot \chi)}$
    \begin{align} \label{eq:hard_modes}
        p=\pm \frac{(S\cdot \chi)\Lambda}{(U\cdot \chi)},\quad \omega=\pm \frac{(S\cdot \chi)^2 \Lambda}{(U\cdot \chi)^2}.
    \end{align}
\end{itemize}

Their classification in \emph{soft} and \emph{hard} refers to their character when approaching the UH. While soft modes remain regular, hard modes diverge when $(U\cdot \chi)=0$, signaling a strong blue-shift of the mode, so that its preferred energy and momentum are independent of the Killing energy $\Omega$.

Notice that there are four solutions close to the UH, while only two exist in the asymptotic region. This signals that the character of the equation \eqref{eq:disp_rel} changes at some intermediate point, which can be obtained by looking at the vanishing of the discriminant of the equation. This happens to occur when
\begin{align}\label{eq:degenerate_cond}
    \frac{(U\cdot \chi)}{(S\cdot \chi)}=1-\frac{3}{2}\left(\frac{\Omega}{\Lambda (S\cdot \chi)}\right)^{\frac{2}{3}}+{\cal O}\left(\Lambda^{-\frac{4}{6}}\right),
\end{align}
which is always in the interior of the Killing horizon. At this point, two of the modes degenerate into one, their value becoming complex beyond this region. This can be seen even more explicitly by solving the equation close to the Killing horizon at leading order in $\Lambda$. There, no coefficients are singular, and the solutions can be simply found to be
\begin{align}\label{eq:modes_KH}
k_\rho =\left\{\frac{\Omega}{(S\cdot \chi)-(U\cdot \chi)},\frac{\Omega}{(S\cdot \chi)+(U\cdot \chi)},\pm \frac{\Lambda}{(U\cdot \chi)}\sqrt{\left[(S\cdot \chi)+(U\cdot \chi)\right]\left[(S\cdot \chi)-(U\cdot \chi)\right]}\right\}.
\end{align}

\begin{figure}
	\centering
	\includegraphics[width=0.5\textwidth]{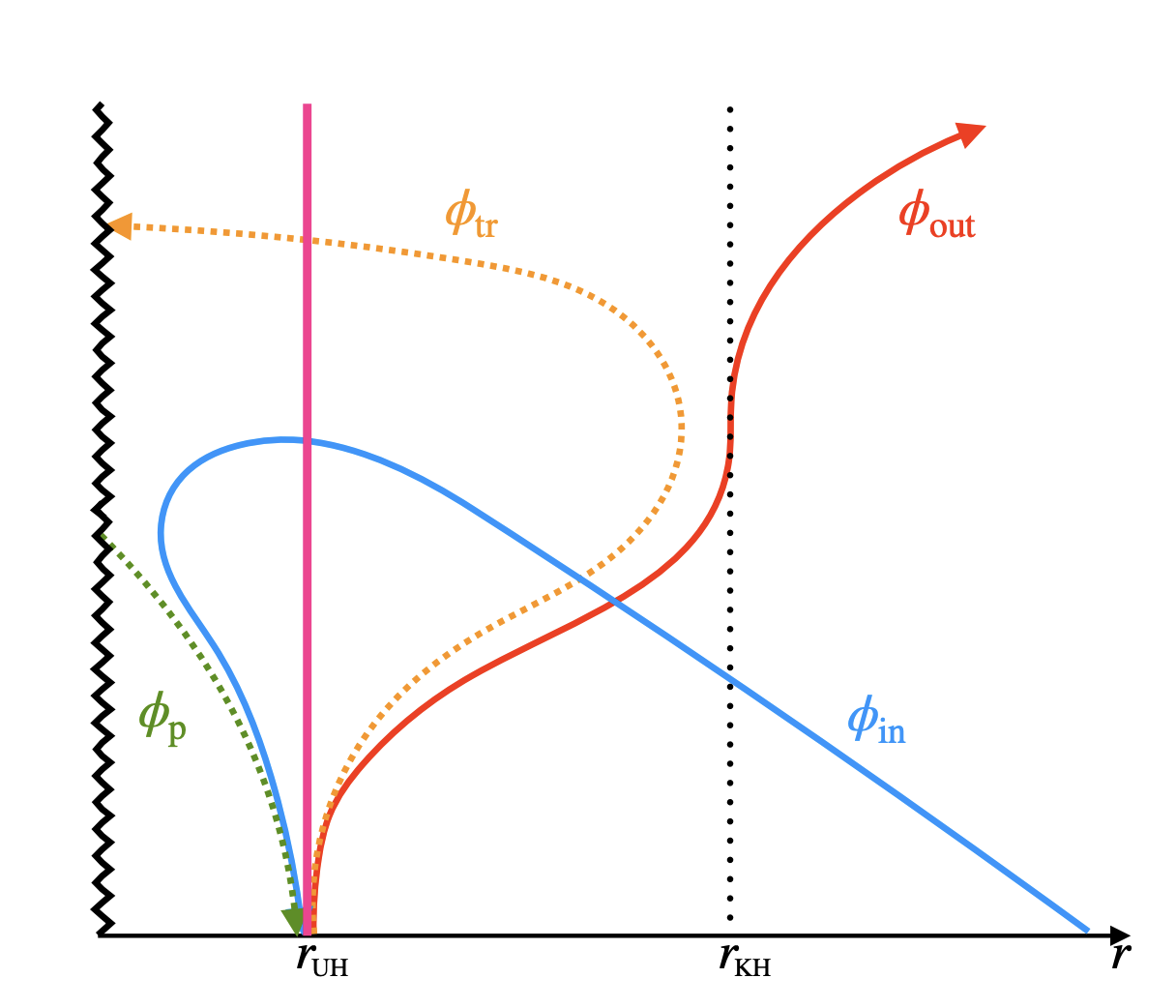}
	\caption{Schematic representation of the modes. The vertical axis denotes proper time, and the direction of propagation of every mode is shown by arrows. The UH is displayed as a solid line, while the Killing horizon is a dotted line. The zigzag line corresponds to the singularity. We see four different modes spanning the radial direction. An outgoing ray $\phi_{\rm out}$, an in-falling one $\phi_{\rm in}$, the trapped complex conjugated mode $\phi_{\rm tr}$, and the partner mode $\phi_{\rm p}$ that lives in the inner region. Figure adapted from \cite{Schneider:2023cuo}.}
        \label{fig:modes}
\end{figure}

We see that again we find in general four solutions. However, two of them degenerate at the surface $(U\cdot \chi)=(S\cdot\chi)$, which corresponds to the Killing horizon and to the leading order of \eqref{eq:degenerate_cond}. They are complex outside of it, and real in its interior. The two other solutions of course mimic the behavior of a relativistic ray at this order \cite{Jacobson:2003vx}. One of them diverges at the Killing horizon due to the coordinate singularity, while the other crosses it without an issue. Note however that the singularity of the former is actually an artifact of perturbation theory. A cartoon picture of the characteristic momentum $k$ is provided in figure \ref{fig:modes}, showing that while the mode lingers for a long time close to the Killing horizon, it never actually diverges there \cite{Cropp:2013sea}. This behavior will have consequences for phenomena related to the emission of Hawking radiation by this kind of black-hole space-times, as we will see in section \ref{sec:hawking}.

The presence of a richer space of solutions behind the Killing horizon is responsible for a plethora of new physical phenomena which are not present in GR, and can potentially lead to observational features. In particular, the presence of the degenerate mode can trigger a super-radiant effect even if the black hole is not rotating \cite{Oshita:2021onq}. In standard GR, superradiance occurs in rotating solutions due to the existence of an ergosphere, enclosed between the Killing and event horizons, where modes with negative Killing energy can go on-shell \cite{Brito:2015oca}. Thanks to this, there are kinematically allowed processes where a body of positive Killing energy $\Omega_1$ enters the ergosphere and splits in two, with one of the halves carrying negative energy and falling into the black hole. If the second half of the body escapes back to the asymptotic region of large radii, it will do it with a killing energy $\Omega_2>\Omega_1$, as a consequence of energy conservation. This Penrose process allows for extraction of energy out of the irreducible mass of the black hole, which is a combination of its mass and angular momentum. If continued in time, it even allows to pump out all the energy stored as angular momentum, leading to a static final state. 

Superradiance provides an explicit way to carry out such energy extraction, by scattering waves onto a rotating black hole in a continuous manner -- typically because of the existence of a confining potential. Whenever the wave number of the incoming wave takes an appropriate range of values, the reflected wave scattered by the black hole carries more energy than the incident one. The presence of superradiance in astrophysical black holes coupled to scalar and vector species with small masses can be used to constrain the value of such mass, by estimating the remaining population of rotating black holes of a given mass and spin, which can be compared with observations \cite{Brito:2015oca}.

The case of Lorentz violating perturbations discussed here can be closely related to this picture. Due to superluminal motion, arbitrary modes can cross the Killing horizon, entering and exiting the region enclosed between it and the UH freely. Within this region, it is possible to excite the degenerate mode, which indeed carries negative Killing energy -- see \cite{Michel:2015rsa,DelPorro:2022kkh,DelPorro:2022vqi}. This possibility was confirmed by numerical evolution of a system akin to \eqref{eq:Lifshitz_action} in \cite{Oshita:2021onq}. There, it is found that whenever $\alpha_4\neq 0$, superradiance can be triggered within certain range of values for the momentum of the scattered wave. This can also be seen by the fact that the potential term entering the Schr\" odinger equation for spherical perturbations on top of the black hole contains a centrifugal term 
\begin{align}
    V_{\rm c}(r)=\frac{l(l+1)U_0^2}{r^2}\left[1+\frac{\alpha_{4}}{\Lambda^2}\left(\frac{l(l+1)}{r^2}-2\frac{U_0^2}{r^2}+\frac{2 \partial_r U_0}{r}\right) \right],
\end{align}
where $U_0$ is the time-component of the aether vector in Schwarzschild time and $l$ the angular momentum eigenvalue. This becomes negative within a narrow region behind the Killing horizon, as discussed in \cite{Oshita:2021onq}. Thus, a reservoir of negative energy, similar to an angular momentum well, becomes accessible for energy extractions. Beyond the pure existence of superradiance, the dynamics of higher derivatives can also influence the ringing of a Lorentz violating black hole formed after a merger. The numerics of \cite{Oshita:2021onq} point towards a faster ringing for $\alpha_{4}>0$, while $\alpha_{4}<0$ leads to long-lived modes.

The results of \cite{Oshita:2021onq} serve as a proof of concept for Lorentz violating superradiance, but several questions remain to be addressed more carefully. Importantly, their methods are only valid within a limited range of couplings and energies, due to the numerical scheme used for integration of the equations of motion. Changing the parameters in the problem quickly leads to violations of the Courant–Friedrichs–Lewy condition \cite{NumRecipes}, thus demanding an alternative approach. In \cite{implicit} an implicit numerical scheme for dispersive equations, based on the well-known Crank-Nicholson method for the heat equation, is developed. Their preliminary results match those in \cite{Oshita:2021onq}, but the numerical method does not suffer from instabilities when large spatial resolutions are required. Moreover, the authors of \cite{implicit} observe a rapid cascade of high energy modes accumulating in the region between the Killing and universal horizons. Since linear perturbations of the scalar mode propagated by Ho\v rava gravity should generically behave as Lifshitz fields -- due to symmetry reasons --, this could indicate a instability of the UH and hence of the solution. However, it is premature to conclude this without a more careful analysis.


\subsection{Hawking radiation from Universal Horizons}\label{sec:hawking}

The fact that black holes evaporate through gravitational emission of particles is, by now, a well established result in GR \cite{Jacobson:2003vx}. It is not only fundamental to the thermodynamical description of black holes, but it has also driven most of our understanding of QFT in curved space-time. In GR, the emission of Hawking radiation is intimately related to the role of the event horizon as a causal boundary, which forbids classical trajectories to exit its envelope. However, quantum tunneling can allow for particles in the vicinity of the potential barrier provided by the horizon to tunnel to the outer region with a non-vanishing probability \cite{Parikh:1999mf}, which follows a black-body law with temperature controlled by the mass $M$ enclosed by the black hole as $T=(8\pi M)^{-1}$. Notice that we are focusing here only on spherically symmetric solutions, where the event horizon corresponds to the Killing horizon of the time-like Killing vector.

In Lorentz violating gravity, the usual event horizon marked by the bending of light-cones is not a causal boundary anymore, since superluminal trajectories can enter and exit it freely. Instead, the UH acts as a trapping surface for all classical trajectories, and it seems reasonable to argue then that all thermal properties, and in particular Hawking radiation, should be recovered in its vicinity. This argument is however too naive and finds many obstructions a priori. Among them we can list the fact that at low energies, the out-going mode that reaches infinity develops a pole at a position perturbatively close to the Killing horizon -- see \eqref{eq:modes_KH}. This means that for those energies, the Killing horizon seems to behave \emph{effectively} as a causal boundary, thus controlling the emission of Hawking radiation through the mechanism briefly described before. This is also supported by computations in analogue gravity, where surfaces which are only approximately Killing horizons lead to particle emission \cite{Unruh:1994je}. On the other hand, there were contradicting results about the potential of the UH for particle emission \cite{Michel:2015rsa}, and even in the cases where the answer was positive, the emission seemed to depend on the particular shape of the dispersion relation for the matter field \cite{Herrero-Valea:2020fqa,Ding:2016srk,Cropp:2016gkn}, a result that can be worrisome, as it allows for the construction of a perpetual mobile of the second kind \cite{Dubovsky:2006vk}.

This conundrum was finally solved recently in \cite{Schneider:2023cuo,Sch2}, where the authors study the behavior of solutions to \eqref{eq:disp_rel} in all regions of space-time. Close to the UH, the naive expectation is actually fullfilled. The hard modes \eqref{eq:hard_modes} peel infinitely at the UH, strictly diverging at its position, meaning that no classical trajectory can escape it. However, quantum tunelling between an inner trajectory and the escaping hard mode can occur with a non-vanishing probability. That happens only because in the inner region, there exist trajectories of negative energy travelling towards the singularity -- green line in Figure \ref{fig:modes} --, which can be seen as trajectories of positive energy trying to escape, providing the inner path for a tunneling particle. In GR this is possible because in the interior of the Killing horizon, particles with negative Killing energy $\Omega$ can go on-shell. Here instead, the reservoir of negative energy is provided by the behavor of the foliation in the neighborhood of the UH, as noted in \cite{DelPorro:2022kkh}.

As previously discussed, the condition of existence of the UH in spherically symmetric solutions corresponds to the vanishing of the lapse $N=0$ in a smooth way -- signalled by $(a\cdot \chi)\propto \partial_r N\neq 0$, where $a_\m=U^\n\nabla_\m U_\n$ is the acceleration of the aether \cite{Bhattacharyya:2015gwa}. Preserving the $\mathbf{C}^2$ character of the solution thus implies that necessarily $N$ must change sign at the UH. This however leads to a causal disconnection of the interior and exterior regions and, more importantly, to a reversal of the flow in preferred time \cite{Bhattacharyya:2015gwa}. This can be seen by performing two subsequent diffeomorphisms, from exterior preferred time $\tau$ to Killing time $t$ -- which is continuous through the UH --, and to interior preferred time $\tilde{\tau}$. The interior lapse $\tilde N$ is then related to the exterior one by
\begin{align}
    \tilde N d\tilde \tau = N d\tau.
\end{align}
Hence, a change of sign in the lapse necessarily implies a change of direction in the flow of preferred time.

Note however that this apparent loss of causality is nevertheless harmless. Since no information can classically escape the UH, it is not possible to build a closed time-loop, preventing the existence of time-machines. Moreover, it is precisely this reversal of time what allows negative energies to exist on-shell behind the UH. As in the case of GR, a particle travelling forward in time with negative energy will correspond to one with positive energy moving backwards in time. This provides the inner trajectory necessary to tunnel a particle out of the UH.

This cartoon picture was made explicit in \cite{DelPorro:2022vqi,Schneider:2023cuo,Sch2}. There, the distribution of particles with action \eqref{eq:Lifshitz_action} emitted by the UH is shown to be thermal with a temperature\footnote{Note that $(a\cdot \chi)<0$ and thus $T>0$.}
\begin{align}\label{eq:T_UH}
    T_{\rm UH}=-\frac{(a\cdot \chi)}{2\pi}.
\end{align}

In previous works in the literature -- see e.g. \cite{Herrero-Valea:2020fqa,Ding:2016srk,Cropp:2016gkn, DelPorro:2022vqi} -- a prefactor carrying dependence on the highest power of momentum in the dispersion relation was included in \eqref{eq:T_UH}. Under its influence, fields endowed with different UV behaviors seemed to lead to different temperatures, which allows to build a perpetual mobile, something certainly dangerous from the point of thermodynamics, unless an extra UV symmetry exists. However, in \cite{Schneider:2023cuo,Sch2}, the authors show that this coefficient was caused by the neglection of a sub-leading term in the solution for $k(\Omega)$, which contributes to particle emission precisely cancelling the controversial coefficient. They also show that no more sub-leading terms contribute to the emission by the UH and thus that the temperature \eqref{eq:T_UH} is universal.

The previous discussion solves part of the problem. The UH radiates with a temperature controlled by its size, in a identical conceptual way to particle emission by quantum fields in GR. However, what does an observer sitting at large radii see? In order to answer this, we must also take into account the propagation of the emitted particles from the UH to the asymptotic region, passing by the Killing Horizon. It is precisely the latter which introduces a modification that leads to a characteristic signature of the presence of Lorentz violations. This is due to the different behavior of wavepackets when approaching the Killing horizon, depending on their energy. 

In order to understand this, let us give a heuristic argument. Note that for very large Killing energy $\Omega$, an arbitrary ray reaching infinity must be travelling at arbitrary large speed, due to the modified dispersion relation. Hence, when crossing the region around the Killing Horizon, it will not feel anything and thus behave as if the horizon were not there at all. The distribution of particles measured by the observer at infinity will thus be thermal for large $\Omega$. On the other hand, when $\Omega$ is small with respect to the Lorentz violating scale in \eqref{eq:Lifshitz_action} -- i.e. $\Omega\ll \Lambda$ -- the ray behaves almost relativistically. It lingers for a long time close to the Killing horizon, reshaping to the form given by the diverging mode in \eqref{eq:modes_KH}. Hence, the observer perceives this ray as perturbatively close to a ray emitted by the Killing horizon. Computing the next-to-leading order correction leads to an \emph{energy dependent} temperature
\begin{align}
    T_{\rm IR}=T_H\left(1+\delta_{c}\frac{\Omega^2}{\Lambda^2}\right),
\end{align}
where $T_H$ is the temperature of the Killing horizon computed in GR, and the factor $\delta_c$ depends on the specifics of the solution. This is akin to what happens in analogue gravity, where no true horizons exist, but instead surfaces where rays linger for a long time act as quasi-particle sources mimicking Hawking radiation \cite{Unruh:1994je}.

At some intermediate energy $\Omega\sim \Lambda$, the previous formula will break down, exiting its range of validity. The full solution must comply with these two limits, while going through a phase transition in the intermediate regime. Nevertheless, its computation is far from the reaching of current techniques. However, a clear conclusion can be extracted from this behavior. The full distribution of particles measured by an observer at large radii will look thermal at large $\Omega$, with temperature $T_{\rm UH}$, while at low $\Omega$ its shape will be controlled by $T_{\rm IR}$. A measure of non-thermality in this very specific way in the spectrum of Hawking radiation emitted by a spherically symmetric black hole is thus a smoking gun of the existence of an underlying Lorentz violating description in the terms discussed here.


\section{Concluding remarks and Roads Ahead}\label{sec:conclusions}

Throughout this text, we have reviewed some of the main results obtained in the last fourteen years of research in Ho\v rava gravity. In particular, we have put emphasis in the developments on the understanding of the core of the proposal -- its renormalizable character. We have shown that the projectable model is indeed fully renormalizable and that it displays candidates for asymptotically free UV fixed points both in 2+1 and 3+1 dimensions. Although the strength of the gravitational coupling flows to arbitrarily large values at low energies, indicating a loss of perturbative control, one should not forget that the theory is otherwise a \emph{bona fide} theory of Quantum Gravity, which could be used to understand some of the puzzles related to the quantum behavior of gravitation in Nature. Moreover, exiting the regime of perturbative control does not mean that the theory is inconsistent, but only that good old perturbative QFT cannot be used to describe the theory within this phase. There is actually a non-vanishing chance that the behavior of pHG at low energies could be akin to QCD, simply requiring non-perturbative techniques to describe its phenomenology. Indeed, pHG seems suitable for application of lattice techniques developed for gauge theories, with many of the problems appearing in the latter -- such as the issue of discretizing fermions -- absent. Additionally, there is the suggestive idea that the scalar mode could lead at low energies to an interaction mimicking the cold dark matter component of our Universe \cite{Mukohyama:2009mz}.

Even though the results in pHG are interesting, the most important question within Ho\v rava gravity is still open and attains the renormalization of the non-projectable model. Due to the presence of second class constraints, the theory cannot be quantized as a gauge theory, in contrast to what happens in pHG. However, we have seen recent progress along this direction. The results of \cite{Bellorin:2022qeu} show that non-local divergences cancel at all order in the loop expansion, which represents the first stone towards building a full proof of renormalizability. Even if such a proof takes time, we are now in such state that computations are feasible, and a sound result is ensured -- either as another stone sustaining the hopes for renormalizabilty, or by finally disproving this possibility.

Beyond its UV behavior, non-projectable Ho\v rava gravity is also interesting because it leads to diverging phenomenology with respect to GR. At low energies there is always a remnant of Lorentz violations, which manifests through the presence of a super-luminal scalar degree of freedom. This leads to observational consequences, which can be used to severely constrain the IR parameter space of the theory, but importantly not to rule it out. Actually, as \cite{Gupta:2021vdj} discusses, Ho\v rava gravity -- or EA gravity in that context -- seems to satisfy observational bounds \emph{as good as GR}. Percolation of Lorentz violations to the SM sector can be very dangerous within this context, since they are strongly constrained \cite{Liberati:2013xla}, but there are mechanisms to avoid them \cite{Pospelov:2010mp}.

Black holes in non-projectable Ho\v rava gravity are also very interesting, since they display a universal trapping surface, named universal horizon, which acts as a causal boundary for trajectories moving with any speed, no matter how large. Although it is unknown whether its existence extends beyond spherical symmetry and staticity, its properties provide a way to extend many of the standard results of GR to the case of Ho\v rava gravity. In particular, it allows for the existence of particle production in a thermal way on its vicinity. Strikingly, an observer sitting far from the black hole would not observe a thermal emission however, since radiation has to cross the Killing horizon sitting in between the observer and the universal horizon, which will deform the spectrum, specially for low energy rays.

Stability of the universal horizon is another open question in the literature, since it requires to account for the effect of higher derivative terms in the action. These are complicated, because they dominate the action at high energies, so one cannot use perturbation theory. Moreover, their presence endows the equations of motion with a dispersive character, which complicates analytical and numerical computations. Very recent results on the propagation of scalar fields with higher derivative kinetic terms -- the Lifshitz scalar field -- show that UV modes tend to cascade and accumulate close to the UH \cite{implicit}, but a proper investigation in the full gravitational case is still beyond the state of the art.

Finally, there is a conundrum on whether a quantum theory of gravitation should cure singularities. In \cite{Lara:2021jul} it is shown that it is not necessarily the case in pHG in 2+1 dimensions, but this work is done under several assumptions and simplifications. The most obvious one is that the values of the couplings in the action are taken to be the same at all scales, disregarding renormalization group flow. Moreover, there could be pathologies associated to space-time dimensionality. Inclusion of the flow, and studying solutions in the mores complicated 3+1 model, could lead to a different conclusion.

As all of this shows, Ho\v rava gravity remains as a very interesting approach to Quantum Gravity in 2023. It is one of the few known theories of gravitation beyond GR that keeps avoiding experimental constraints, fulfilling them with an astonishing agreement. Moreover, and even though its structure of interactions is complicated, its UV behavior is very rich, and there is an increasing possibility that the full theory is renormalizable. An even if this happens to not to be the case, the projectable model \emph{is} renormalizable, and can therefore be used at the very least as a toy model to understand several aspects of gravitation. A lot of work remains to be done, but the goal of finding a UV completion of GR seems to be closer than ever.


\section*{Acknowledgements}
I am deeply grateful to my collaborators Enrico Barausse, Andrei Barvinsky, Miguel Bezares, Diego Blas, Neil Cornish, Nicola Franchini, Francesco del Porro, Toral Gupta, Aron Kov\'acs, Guillermo Lara, Stefano Liberati, Marcelo Rubio, Raquel Santos-García, Marc Schneider, Sergey Sibiryakov, Christian Steinwachs, Kent Yagi, and Nicolás Yunes; whose contributions to past and current projects are fundamental to many of the results discussed here. I also want to thank all the participants in the Focus Week Program: Lorentz violations in Gravity, held at the Institute for the Fundamental Physics of the Universe (IFPU) in Trieste (Italy), in July 2023. Finally, let me thank the editors of the EPJP special issue on “Higher Derivatives in Quantum Gravity" for inviting me to contribute with this review. My work has been supported by the Spanish State Research Agency MCIN/AEI/10.13039/501100011033 and the EU NextGenerationEU/PRTR funds, under grant IJC2020-045126-I; and by the Departament de Recerca i Universitats de la Generalitat de Catalunya, Grant No 2021 SGR 00649. IFAE is partially funded by the CERCA program of the Generalitat de Catalunya.

\bibliography{biblio}

\end{document}